\documentclass[aps,prb,twocolumn,superscriptaddress,groupedaddress]{revtex4}

\usepackage{graphicx}
\usepackage[export]{adjustbox}
\usepackage{dcolumn}
\usepackage{bm}
\usepackage{hyperref}
\usepackage[mathlines]{lineno}
\usepackage{amsfonts}
\usepackage{amssymb}
\usepackage{tikz}
\usepackage{mathtools}

\newlength{\thelinewidth}
\thelinewidth=\textwidth

\hypersetup{colorlinks=true, linkcolor=blue, citecolor=blue}

\begin{document}
\title{Coherent coupling between multiple ferrimagnetic spheres \\and a microwave cavity at millikelvin temperatures}
\author{N.~Crescini}
	\email{nicolo.crescini@phd.unipd.it}
	\affiliation{INFN-Laboratori Nazionali di Legnaro, Viale dell'Universit\`a 2, 35020 Legnaro (PD), Italy}
	\affiliation{Dipartimento di Fisica e Astronomia, Via Marzolo 8, 35131 Padova, Italy}
\author{C.~Braggio}
	\affiliation{Dipartimento di Fisica e Astronomia, Via Marzolo 8, 35131 Padova, Italy}
	\affiliation{INFN-Sezione di Padova, Via Marzolo 8, 35131 Padova, Italy}
	
\author{G.~Carugno}
	\affiliation{Dipartimento di Fisica e Astronomia, Via Marzolo 8, 35131 Padova, Italy}
	\affiliation{INFN-Sezione di Padova, Via Marzolo 8, 35131 Padova, Italy}
\author{A.~Ortolan}
	\affiliation{INFN-Laboratori Nazionali di Legnaro, Viale dell'Universit\`a 2, 35020 Legnaro (PD), Italy}
\author{G.~Ruoso}
	\affiliation{INFN-Laboratori Nazionali di Legnaro, Viale dell'Universit\`a 2, 35020 Legnaro (PD), Italy}
	
\date{\today}

\begin{abstract}
The spin resonance of electrons can be coupled to a microwave cavity mode to obtain a photon-magnon hybrid system. These quantum systems are widely studied for both fundamental physics and technological quantum applications. In this article, the behavior of a large number of ferrimagnetic spheres coupled to a single cavity is put under test.
We use second-quantization modeling of harmonic oscillators to theoretically describe our experimental setup and understand the influence of several parameters.
The magnon-polariton dispersion relation is used to characterize the system, with a particular focus on the vacuum Rabi mode splitting due to multiple spheres.
We combine the results obtained with simple hybrid systems to analyze the behavior of a more complex one, and show that it can be devised in such a way to minimize the degrees of freedom needed to completely describe it. By studying single-sphere coupling two possible size-effects related to the sample diameter have been identified, while multiple-spheres configurations reveal how to upscale the system.
This characterization is useful for the implementation of an axion-to-electromagnetic field transducer in a ferromagnetic haloscope for dark matter searches. Our dedicated setup, consisting in ten 2\,mm-diameter YIG spheres coupled to a copper microwave cavity, is used for this aim and studied at mK temperatures.
Moreover, we show that novel applications of optimally-controlled hybrid systems can be foreseen for setups embedding a large number of samples.
\end{abstract}

\maketitle


	\section{Introduction}
	\label{sec:intro}
In the fruitful and renowned field of light-matter interaction\cite{haroche}, the study of hybrid quantum systems based on magnonics gave outstanding results \cite{Clerk2020,Lachance_Quirion_2019}.
Magnons, quanta of spin excitations, can coherently couple to photons through a magnetic dipole interaction\cite{Kittel2004,PhysRev.143.372}. The electron spin resonance of a magnetic material can thus be coupled to the rf magnetic field of a microwave cavity mode, to form a photon-magnon hybrid system (HS)\cite{PhysRevLett.111.127003,PhysRevLett.113.083603,PhysRevApplied.2.054002,Zhang2015} at GHz frequencies.
This kind of HSs are employed in fields including, but not limited to, the development of quantum computers\cite{Ladd2010}, quantum networks\cite{Kimble2008,RevModPhys.87.1379}, and quantum sensing\cite{RevModPhys.89.035002}.
In these areas, the coherent coupling of spin ensembles, microwave cavities\cite{PhysRevLett.111.127003,PhysRevLett.113.083603,PhysRevLett.113.156401,PhysRevApplied.2.054002,Zhang2015,PhysRevLett.114.226402,PhysRevLett.105.140501} and superconducting quantum circuits\cite{Tabuchi405,wang2019quantum,TABUCHI2016729} is used to create quantum memories \cite{doi:10.1080/09500340.2016.1148212,Kurizki3866}, to convert optical photons to microwaves\cite{cate} and vice versa\cite{PhysRevA.92.062313,PhysRevLett.113.203601,PhysRevB.93.174427}, and to detect single quanta of magnetization\cite{Lachance-Quirion425}. In spintronics, magnons\cite{Chumak2015} are suitable carriers of spin information, and have several advantages over electrical currents used in modern electronics.
Furthermore, the investigation of non-Hermitian quantum mechanics\cite{moiseyev2011non} is currently pursued with HSs.
Exceptional points are the signature of non-Hermitian physics\cite{Bender_2007,Heiss_2012}, and their observation was recently proposed\cite{PhysRevX.6.021007}, and experimentally verified\cite{Zhang2017} in photon-magnon HSs. These results were followed by the demonstration of exceptional surfaces\cite{PhysRevLett.123.237202}, and the measure of a third order exceptional point has been proposed in systems of multiple ferrimagnetic spheres\cite{PhysRevB.99.054404}.
Besides the study of non-Hermitian quantum mechanics, different applications can be envisioned for these points, for example as sensitive probes of magnetic fields\cite{PhysRevB.99.214415}.
These and many other studies\cite{PhysRevLett.124.053602,PhysRevLett.120.057202,PhysRevLett.123.127202,PhysRevA.93.021803,yigbell} maintain that photon-magnon HSs are an outstanding testbed for the study of many physical phenomena.

The study of HSs comprising a large amount of magnetic samples has been explored to create gradient memories\cite{Zhang2015}, or to detect faint pseudo-magnetic fields\cite{quaxepjc}.
In precision physics, rf spin-magnetometers based on HSs are used as axion haloscopes under the name of ferromagnetic haloscopes\cite{quaxepjc,FLOWER2019100306} (FH). In these devices, a large number of spins increases the sensitivity to magnetic oscillations induced by dark matter axions. As the axion couples to magnons\cite{BARBIERI1989357} in a FH the HS acts as an axionic-to-electromagnetic field transducer.

This work reports on the coherent coupling of ten 2.1\,mm-diameter ferrimagnetic spheres to a cylindrical copper cavity, with the aim of implementing the resulting HS in a FH.
Our results, however, go beyond this sole purpose, and investigate the dynamics of such large and complex systems, which are promising for multiple usages\cite{Zhang2015,PhysRevA.93.021803,PhysRevB.97.014419,PhysRevB.99.094407,PhysRevResearch.1.023021,PhysRevLett.121.087204}.
In our setup, the strong coupling regime is largely reached already with a single sphere, thus we can study the scaling without qualitatively changing the behavior of the HS. We singularly tested spheres with diameters ranging from 0.5\,mm to 2.5\,mm and observed two different size-effects, affecting the $g$-factor and the sphere's zero-field splitting. By choosing a diameter of 2.1\,mm, we focus on the coherent coupling of multiple spheres.
The multi-samples design is preferred to a single rod of magnetic material for practical reasons, like the material availability and geometric demagnetization.
The influence of the dimension and relative distance of the samples are consistently accounted for in a model which precisely reproduces the experimental data.
Our study thus demonstrates that it is feasible to scale up a HS while preserving its optimal control.

In Sec.\,\ref{sec:sqm} we detail the model used to describe the experiment. Sec.\,\ref{sec:er} illustrates the setup, and is further divided into two parts: the first (\ref{sec:era}) explains the room temperature tests used to understand the behavior of single spheres, while the second (\ref{sec:erb}) reports on the coupling of many multiple spheres to the cavity, and on the room temperature and ultra cryogenic tests of the resulting HS.
The last part of the paper, Sec.\,\ref{sec:fdc}, is dedicated to the possible improvements of the setup, in terms of increasing the quantity of material coupled to the cavity, and to a summary of the obtained results.


	\section{Second-quantization model}
	\label{sec:sqm}
This section deals with the model used to study the HS dynamics, which essentially consists in a system of coupled harmonic oscillators in the second-quantization formalism\cite{walls2007quantum,scully1999quantum} that efficiently describes the number of (quasi-)particle in a state by using creation and annihilation operators to add or remove a quanta. Similar approaches have been used to study the interaction of multiple qubits in a cavity\cite{PhysRevLett.103.083601}, and test the Tavis-Cummings model\cite{PhysRev.170.379} for a low number of two-level systems. Since we are interested in describing resonant quanta, raising and lowering operators share the same algebra of a collection of interacting quantum harmonic oscillators.
Let us consider $N$ photon modes $X=\{ x_1, x_2, \dots, x_N \}$, and $M$ magnon modes $Y=\{ y_1, y_2, \dots, y_M \}$, and label with $\omega$ and $g$ their frequencies and couplings, respectively. Every mode $X$ can couple to any of the modes $Y$, so the Hamiltonian of the system in the rotating wave approximation reads
\begin{align}
	\begin{split}
	H =	&\hbar\sum_{x\in X} \omega_x \ x^\dag x + \hbar\sum_{y\in Y} \omega_y y^\dag y \\
		&+\sum_{x\in X}\sum_{y\in Y}g_{xy}(x^\dag y+y^\dag x) \\
		&+ \sum_{i\neq j}g_{y_iy_j}(y^\dag_i y_j + y_j^\dag y_i),
	\label{eq:H}
	\end{split}
	\end{align}
where $\hbar$ is the reduced Planck constant, and $x^\dag$, $y^\dag$ ($x$, $y$) are the creation (annihilation) operators of the corresponding mode quanta.
Our interest lies in the evolution of the mean values of $x$ and $y$ operators that can be calculated by Heisenberg-Langevin equations. The effects of dissipations in a resonant system due to its coupling with a thermal reservoir can be easily taken into account by adding an imaginary part to the mode frequencies which correspond to their linewidths $\gamma$s.
Thus the equations of motion for the HS evolution can be recast as a first order system of differential equations (see Appendix \ref{app:matrix} for the complete derivation), and its solution can be readily found in Fourier space. The associated HS matrix reads
\begin{widetext}
	\begin{equation}
	{\cal H} =
	\begin{pmatrix}
	
		\begin{array}{ c c c c }
		 \omega_{x_1}-i\gamma_{x_1}/2 	& 0 							& \dots				& 0 \\
		 0 						& \omega_{x_2}-i\gamma_{x_2}/2 		& \dots				& 0	\\
		 \vdots 					& \vdots						& $\quad$\ddots$\quad$		& \vdots \\
		0						& 0 							& \dots				& \omega_{x_{_N}}-i\gamma_{x_{_N}}/2
		\end{array}
	&
	\hspace{-.75cm}
		\begin{array}{| c c c c}
		g_{x_1y_1} 					& g_{x_1y_2}					& \dots				& g_{x_1y_{_M}} \\
		g_{x_2y_1} 					& g_{x_2y_2} 					& \dots				& g_{x_2y_{_M}} \\
		$\qquad$\vdots$\qquad$			& $\qquad$\vdots$\qquad$			& $\quad$\ddots$\quad$		& $\qquad$\vdots$\qquad$ \\
		g_{x_{_N}y_1}				& g_{x_{_N}y_2}					& \dots				& g_{x_{_N}y_{_M}}
		\end{array}\\ \ \\
	\hline \\
		\begin{array}{ c c c c |}
		g_{x_1y_1} 					& g_{x_2y_1}					& \dots				& g_{x_{_N}y_1} \\
		g_{x_1y_2} 					& g_{x_2y_2} 					& \dots				& g_{x_{_N}y_2} \\
		$\quad$\vdots$\qquad$			& $\qquad$\vdots$\qquad$			&$\quad$\ddots$\quad$		& $\qquad$\vdots$\qquad$ \\
		g_{x_1y_{_M}}				& g_{x_2y_{_M}}					& \dots				& g_{x_{_N}y_{_M}}
		\end{array}
	&
	\hspace{-1cm}
		\begin{array}{ c c c c }
		 \omega_{y_1}-i\gamma_{y_1}/2 	& g_{y_2y_1}					& \dots				& g_{y_1y_{_M}} \\
		 g_{y_1y_2}				& \omega_{y_2}-i\gamma_{y_2}/2 		& \dots				& g_{y_2y_{_M}} \\
		 \vdots 					& \vdots						& $\quad$\ddots$\quad$		& \vdots \\
		 g_{y_1y_{_M}} 				& g_{y_2y_{_M}}				& \dots				& \omega_{y_{_M}}-i\gamma_{y_{_M}}/2
		\end{array}
		
	\end{pmatrix}.
	\vspace{.25cm}
	\label{eq:Heff}
	\end{equation}
\end{widetext}
The top-left block stands for the cavity modes, which are considered uncoupled to each other, as typically their resonant frequency difference is much larger than their linewidths. The lower-right block shows the material magnetic modes, which may auto-interact mainly thanks to the dipole coupling between different spheres (see Section \ref{sec:er}). The off-diagonal blocks are the couplings between the cavity and magnetic modes.
The matrix in Eq.\,(\ref{eq:Heff}) completely describes the system dynamics, and can be used to calculate the dispersion relation of the magnon-polariton.

In our study we variate the static magnetic field to understand how it affects the transmission function of the HS. The spin precession frequency of a free electron in a generic magnetic field $B$ is $\omega_0(B)=\gamma B$, with $\gamma=(2\pi)28\,\mathrm{GHz/T}$.
This basic element can be used in the theoretical model to account for a variation of the magnetostatic mode frequencies $\omega_Y$. In particular, we write the frequencies of the $Y$ modes as
	\begin{equation}
	\omega_Y(B)=\gamma (B+ b_Y),
	\label{eq:omegam}
	\end{equation}
where the zero field shift $\gamma b_Y$ is a constant depending on the magnetostatic mode under consideration. In Eq.\,(\ref{eq:omegam}) we choose a linear dependence between $\omega$ and the applied static field, but it is straightforward to include non-linear magnetic field dependencies if any are present.
The poles of the cavity magnon-polariton dispersion relation can be found by solving the characteristic equation $\det(K(\omega,\omega_Y) )=0$, where
	\begin{equation}
	K(\omega,\omega_Y) = \omega \mathbb{I}_{N+M}-{\cal H}(\omega_Y),
	\label{eq:k}
	\end{equation}
and $\mathbb{I}_{N+M}$ is the identity matrix of dimension $N+M$. Here $\omega$ can be seen as the frequency of a monochromatic probe signal.
Experimental spectra are collected through two antennas coupled to the resonant cavity. In principle, we should consider the transmission of all the cavity modes (see Appendix\,\ref{app:matrix}), but, as the model does not account for the antenna coupling strength, we can restrict to one mode and neglect other transmission channels.
To calculate the transmission spectra of the HS, we can consider the monochromatic system excitation coupled to the first cavity mode $x_1$
	\begin{equation}
	s_{x_1}(\omega,\omega_Y) = \frac{\gamma_{x_1}\omega^2}{2}|(K^{-1}(\omega,\omega_Y)\cdot\hat{e}_1)_1|^2,
	\label{eq:s}
	\end{equation}
where $\hat{e}_1$ is the unit vector of $x_1$. Note that the parameters of $\cal H$ can be determined experimentally by fitting the transmission spectra to Eq. (\ref{eq:s}).

In the following, to indicate a particular HS configuration we will write the Hamiltonian of the system, and take for granted that to get the dispersion relation one follows Eq.\,(\ref{eq:k}) and (\ref{eq:s}).

To reduce the complexity of the HS, and understand how the different couplings affect its dynamics, we reduce the number of considered modes to four: two cavity modes $c$ and $d$, and two magnetic modes $m$ and $n$. The mode $c\equiv x_1$ corresponds to the one coupled to the antennas, and $m\equiv y_1$ is the Kittel mode with frequency $\omega_m$. The resulting Hamiltonian is
	\begin{equation}
	{\cal H}_4=
	\begin{pmatrix}
	\omega_c - \frac{i}{2}\gamma_c	&	0				&	g_{cm}					&	g_{cn}			\\
	0				&	\omega_d -\frac{i}{2}\gamma_d	&	g_{dm}					&	g_{dn}			\\
	g_{cm}			&	g_{dm}				&	\omega_m - \frac{i}{2}\gamma_m	&	g_{mn}			\\
	g_{cn}			&	g_{dn}				&	g_{mn}					&	\omega_n -\frac{ i}{2}\gamma_n	\\
	\end{pmatrix}.
	\label{eq:Hcdmn}
	\end{equation}
	
\begin{figure}
	\begin{tikzpicture}
	\node at (-2.25,2) {\includegraphics[width=.23\textwidth]{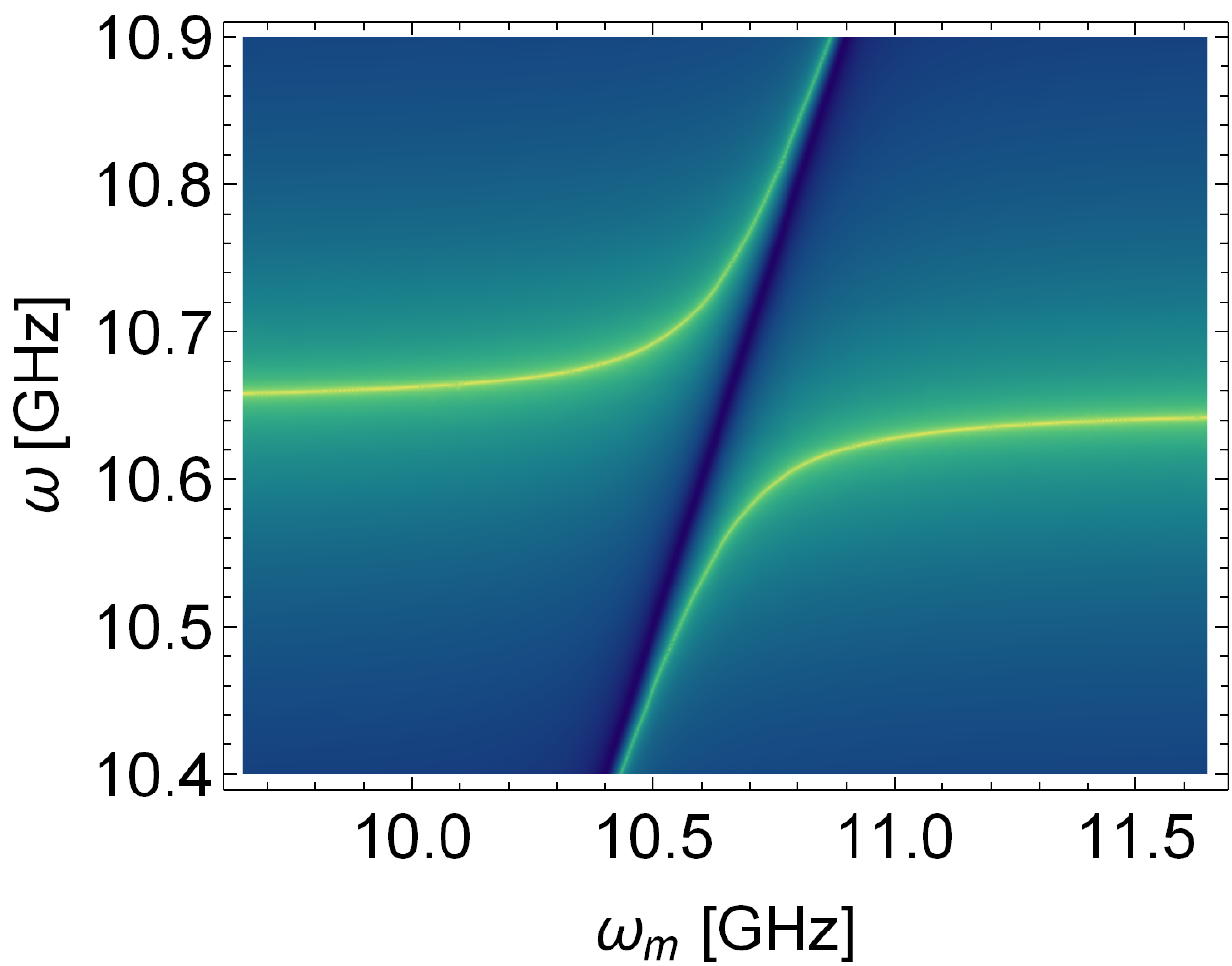}}; %
	\node at (2,2) {\includegraphics[width=.23\textwidth]{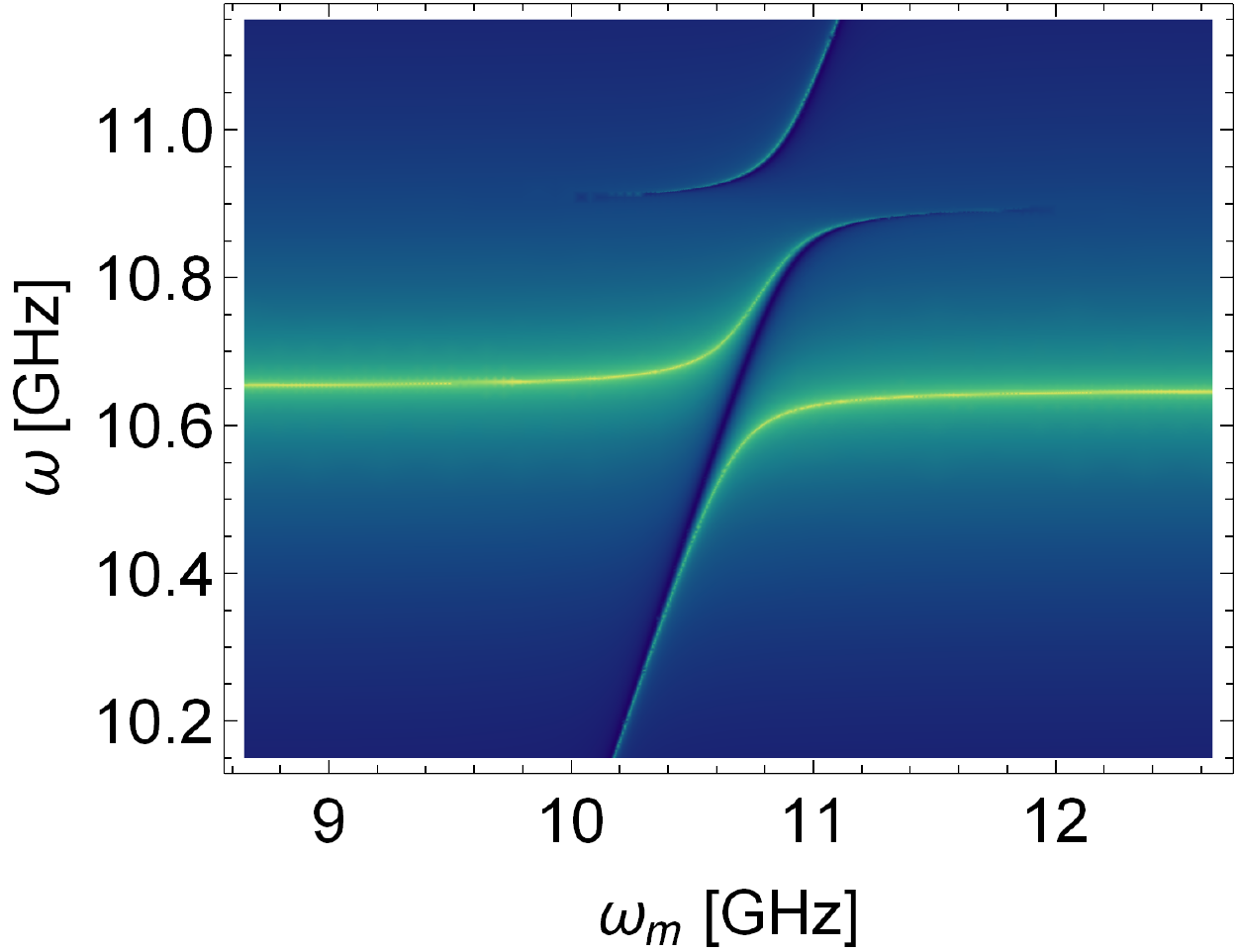}}; 
	\node at (-2.25,-1.5) {\includegraphics[width=.23\textwidth]{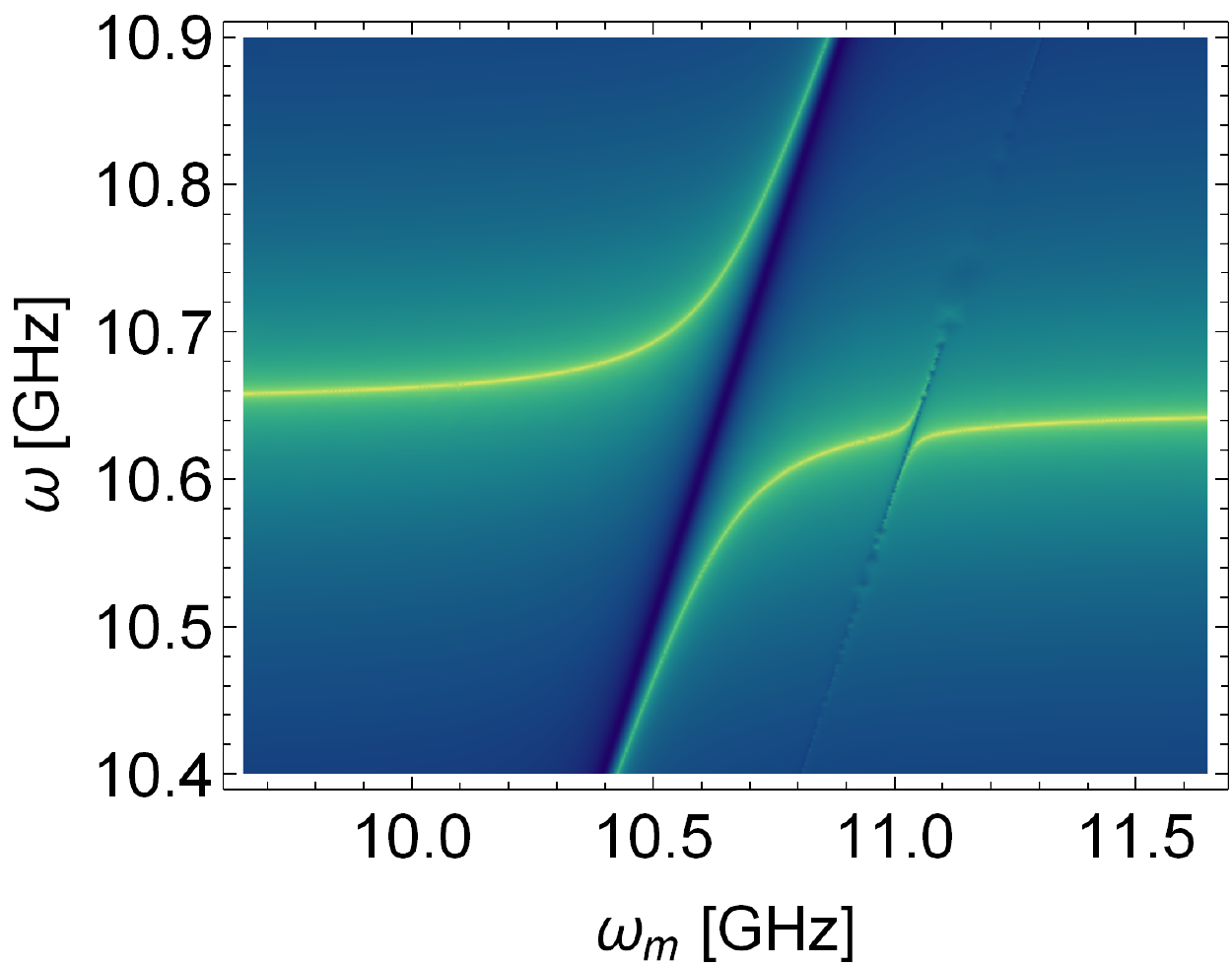}}; 
	\node at (2,-1.5) {\includegraphics[width=.23\textwidth]{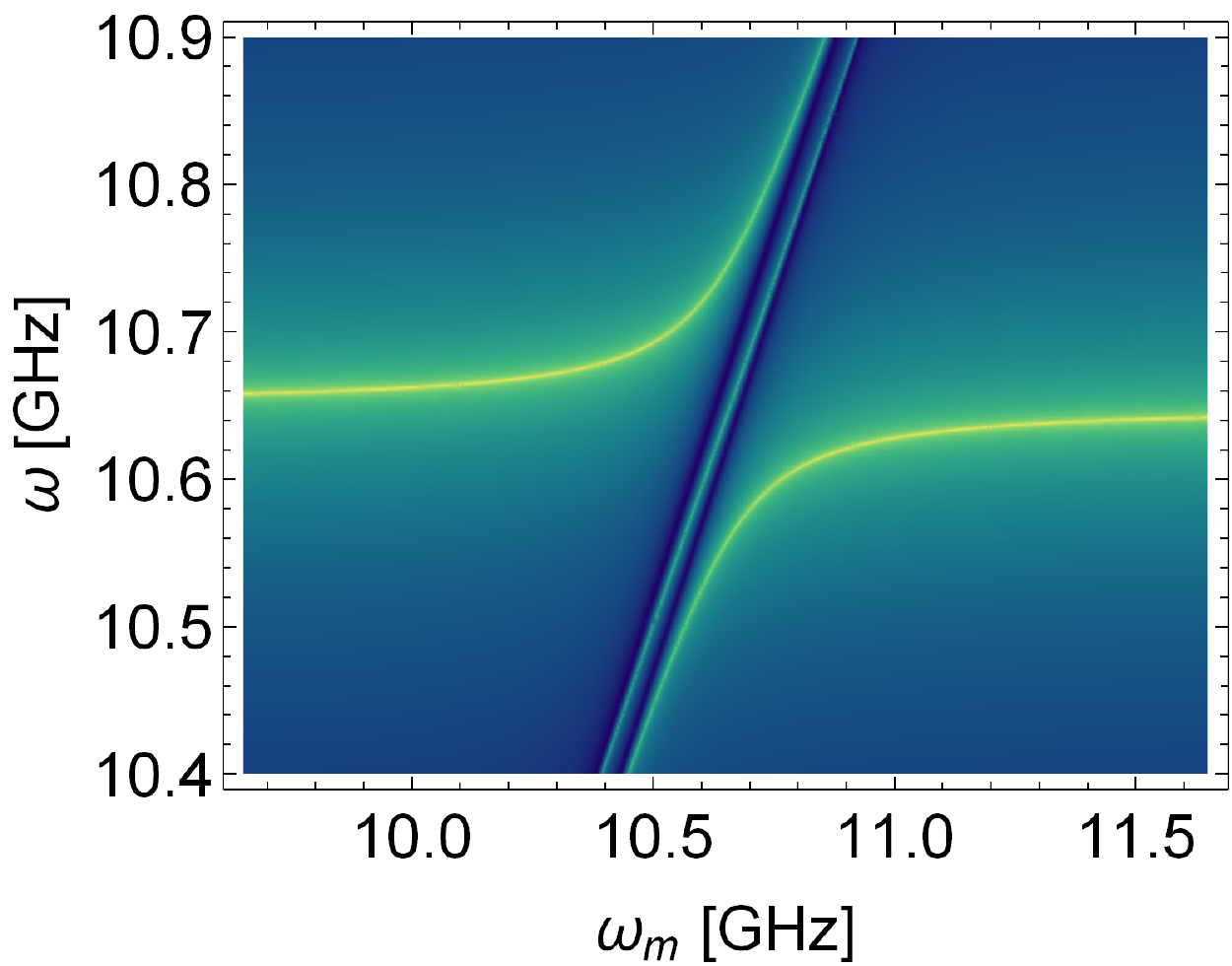}}; 
		\node at(-3,3) { \textcolor{white}{(a)} };
		\node at(1.25,3) { \textcolor{white}{(b)} };
		\node at(-3,-.5) { \textcolor{white}{(c)} };
		\node at(1.25,-.5) { \textcolor{white}{(d)} };
		\node at(-1.25,1.25) { \textcolor{white}{$c$-$m$} };
		\node at(3,1.25) { \textcolor{white}{$c$-$m$-$d$} };
		\node at(-1.25,-2.25) { \textcolor{white}{$n$-$c$-$m$} };
		\node at(3,-2.25) { \textcolor{white}{$c$-$m$-$n$} };		
	\end{tikzpicture}
	\includegraphics[width=.45\textwidth]{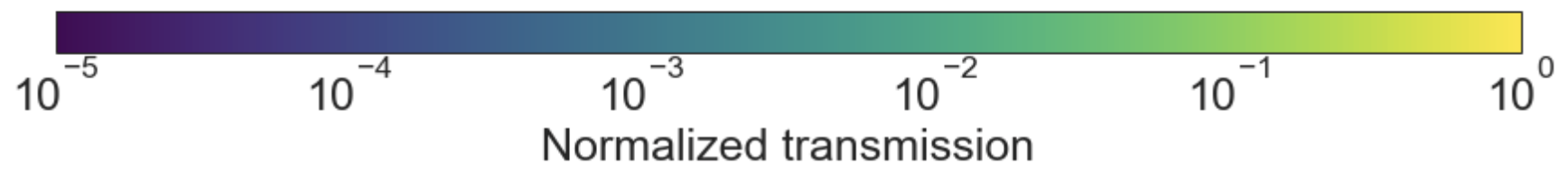}
	\caption{Plots of the transmission spectra $s_{x_1}(\omega,\omega_0)$ illustrating the dispersion relations of the magnon-polariton systems. The color scale is in logarithmic arbitrary units, where blue corresponds to low and light green to high transmission. The dash between the labels of two modes indicates that the two are coupled. Upper plots: normal anticrossing curve (a), and coupling of the Kittel mode with the two cavity modes $c$ and $d$ (b). Lower plots: (c) $c$-mode coupled to two magnetostatic modes $m$ and $n$, with $b_n\neq0$; (d) $c$-mode coupled to $m$, coupled to $n$, with $b_n=0$ and $g_{mn}=25$ MHz. One can obtain the same plot with $g_{mn}=0$ and $b_n\sim0$. See text for further details.}
	\label{fig:cdmn}
	\end{figure}
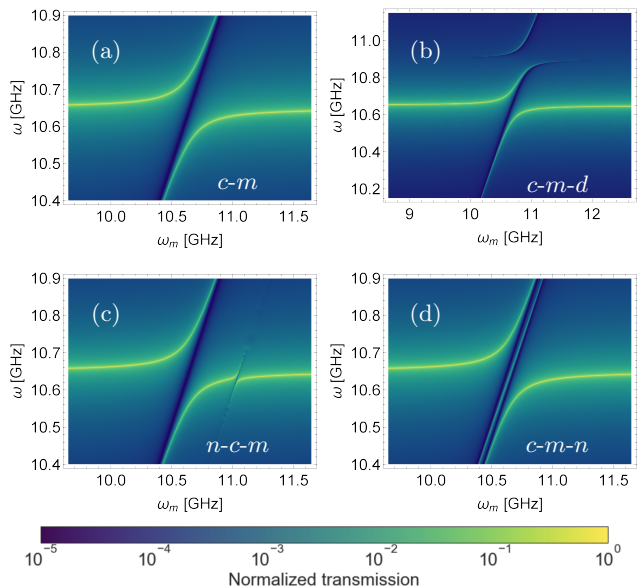

The resonant frequency of the cavity modes are set to $\omega_c=(2\pi)10.65\,$GHz and $\omega_d=(2\pi)10.9\,$GHz, and the linewidths to $\gamma_c=\gamma_d=(2\pi)1.0\,$MHz, while the linewidths of the magnon modes $\gamma_m=\gamma_n=(2\pi)2.0\,$MHz.
The effects of the different terms are studied by first setting to zero all the couplings but $g_{cm}$, set to $(2\pi)$90\,MHz to be close to the value measured with a single 2.1\,mm diameter sphere.
Then, the other couplings are selectively turned on one by one, and the resulting dispersion relations are compared in Fig.\,\ref{fig:cdmn}.
	
Plot \ref{fig:cdmn}(a) is a usual anticrossing curve, where the coupling $2g_{cm}$ is the frequency splitting of the two HS modes when $\omega=\omega_m$. This HS is the starting point for all the following considerations.

The interaction of the Kittel magnetic mode $m$ with two cavity modes $c,d$ is considered in Fig.\,\ref{fig:cdmn}(b), where we set equal couplings $g_{cm}=g_{dm}$, producing a combination of two anticrossing curves. 
The second cavity mode $d$, at the frequency $\omega_d=(2\pi)10.9\,$GHz, is not coupled to the antennas, as is shown by Eq.\,(\ref{eq:s}). For this reason no transmission happens at $\omega_d$ if not mediated by the $m$ mode.

Fig.\,\ref{fig:cdmn}(c) shows a single cavity mode $c$ coupled to two independent magnetostatic modes, $m$ and $n$. The frequency of the mode $n$ is obtained by shifting $\omega_m$ of $\gamma b_n=(2\pi)0.4\,$GHz [see Eq.\,(\ref{eq:omegam})], corresponding to a field bias of about $15\,$mT.
This offset field may describe higher order modes\cite{Princep2017,PhysRevB.101.014439} or two spheres coupled to the same mode but through slightly different fields.
 In fact, as we shall see in the following Section, a size effect related to the diameter of the spheres \cite{PhysRev.105.390,PhysRev.81.477,doi:10.1063/1.2185878,doi:10.1139/p58-114,Krupka_2018} is the presence of a bias field. The coupling of this mode to the cavity mode is arbitrarily set to $g_{cn}=(2\pi)25\,$MHz to roughly resemble the coupling to higher order modes\cite{PhysRevB.101.014439}.

The dispersion relation (d) of Fig.\,\ref{fig:cdmn} is obtained by adding a magnon-magnon coupling $g_{mn}=(2\pi)25\,$MHz between the magnetostatic mode $n$ and the Kittel mode $m$, while keeping $n$ uncoupled from the cavity mode $c$. A third hybrid mode appears, its resonance frequency lies between the two ones of the $c$-$m$ system, and dispersively shifts them.
It is interesting to note that a similar result can be obtained by using a small detuning $b_n$ in the previous $n$-$c$-$m$ configuration but with no $g_{mn}$ coupling and such that $\gamma_m \ll \gamma b_n < 2\sqrt{g_{cm}^2+ g_{cn}^2}$. This means that an identical dispersion relation results from hybridizing, under the same static field, a cavity and two spheres with a different offset field $b_Y$.
The central mode is usually referred to as dark mode, which was studied e.\,g. for qubits\cite{PhysRevLett.103.083601} and for magnons\cite{Zhang2015,doi:10.1063/1.5121618}.
The fact that the effect of a magnon-magnon coupling results similar to the one of a $n$-$c$-$m$ system with a small bias $b_n$ is further discussed and experimentally explored in Sec.\,\ref{sec:erb}.

When $M$ magnon modes are degenerate (i.\,e. have the same resonant frequency) they couple with a cavity mode as a single oscillator, and the vacuum Rabi splitting, the frequency difference of the two hybrid modes, scales as the root of the sum of the squared couplings.
In this case the two-modes system $c$-$m$ holds.
Ideally, the addition of more spheres gives the same result, and the splitting scales as the square root of the total number of spins.
However, a large increment of the number of spheres in a single cavity forces us to face the dynamics outlined in Fig.\,\ref{fig:cdmn}.
These effects can be combined to obtain more convoluted dispersion relations, which can be explained thanks to the understanding of their basic elements provided by this Section.


	\section{Experimental results}
	\label{sec:er}
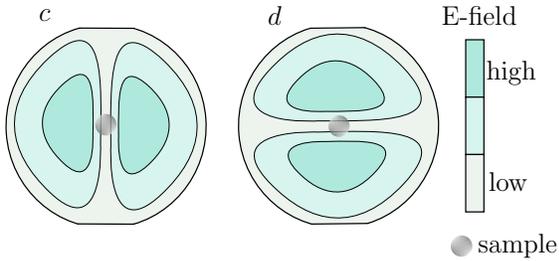
\begin{figure}
	\centering
	\begin{tikzpicture}
	\scalebox{.85}{
	\node at(0,0) {\includegraphics[width=.5\textwidth]{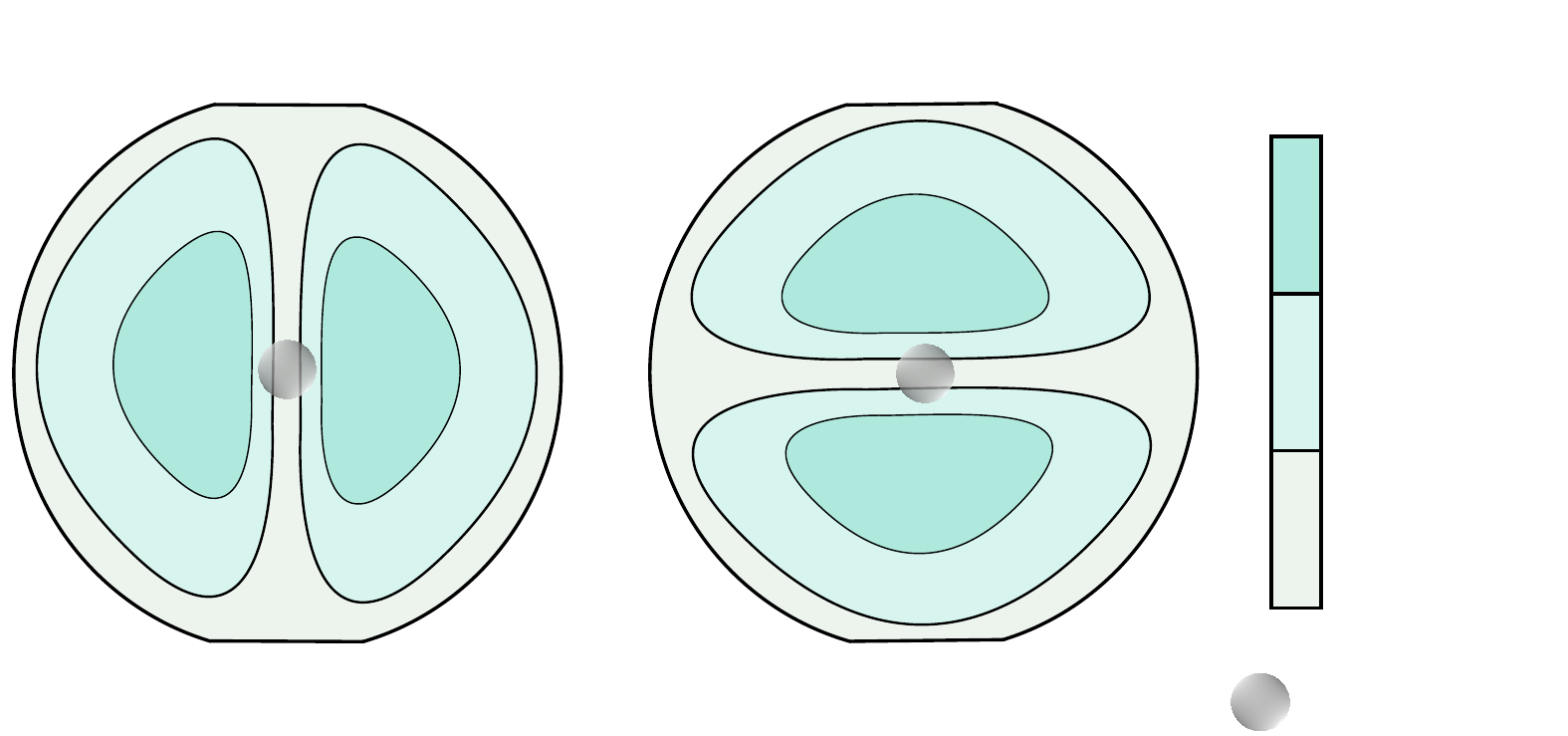}};
	\node	at(3.6,-1.9) {\large sample};
	\node	at(3,1.7) {\large E-field};
	\node	at(3.5,0.8) {\large high};
	\node	at(3.45,-0.9) {\large low};
	\node	at(-3.8,1.7) {\large $c$};
	\node	at(-0.2,1.7) {\large $d$};
	}
	\end{tikzpicture}
	\caption{Section of the cavity body geometrical shape, and profile of the $c$ and $d$ modes. The sample lies on a maximum of the rf magnetic field for both modes.}
	\label{fig:tm110}
	\end{figure}

The photonic resonators of our setup are copper cavities with a cylindrical body and cone-shaped end caps\cite{PhysRevD.99.101101}. The TM110 mode of a perfectly cylindrical cavity is degenerate for rotations about its axis, complicating the coupling to the sample. To lift the degeneracy, the symmetry of the cavity body is broken by two flat walls which reduce its diameter, as shown in Fig.\,\ref{fig:tm110}. As a consequence, the two polarizations have frequencies which differs about 200\,MHz, and
we identify the two corresponding modes with the $c$ and $d$ appearing in Eq.\,(\ref{eq:Hcdmn}).
The magnetic material is Yttrium Iron Garnet (Y$_3$O$_5$Fe$_{12}$), a ferrimagnetic insulator\cite{CHEREPANOV199381,doi:10.1063/1.1723117,doi:10.1063/1.1735216,princepyig} widely known due to its applications in microwave electronics \cite{Pozar:882338}, mainly related to its narrow linewidth\cite{PhysRev.110.1311,PhysRevLett.3.32,yigline}.
The values of resonance frequencies, linewidths and couplings used in the previous Section were chosen to resemble the measured ones of copper cavities and of YIG.

We tested several spheres of different diameter to infer the maximum size that can be used and characterized without taking into account perturbative effects.
Except for 0.5\,mm and 1\,mm-diameter spheres, the others were manufactured on site from a large YIG single crystal. After cutting the crystal into small cubes, spheres are obtained by grinding them with the procedure described in Appendix \ref{app:tumbling}. In the measurement setup, a fused silica pipe and PTFE cups hold the spheres in a position close to the cavity center and let them rotate freely, guaranteeing alignment of their easy axis to the external magnetic field. Such condition is necessary in order to coherently maximize the coupling strength as prescribed in [\onlinecite{quaxepjc}]. In the following tests this condition is always fulfilled, and in Sections \ref{sec:era} and \ref{sec:erb} it is analyzed and understood.

The magnetic field is provided by a superconducting magnet, and care is taken to place the mounted spheres at its center: within a cylindrical region of 0.5\,cm diameter and 7\,cm height the homogeneity of the field is better than 7\,ppm, guaranteeing no inhomogeneous broadening of the electron spin resonance.
The linearity and absence of offsets for the relation between the power supply current and the magnetic field has been verified using a paramagnetic BDPA sample, whose $g$-factor is remarkably close to the one of the free electron. Further details on the experimental apparatus can be found in Appendix \ref{app:apparatus}.


The aim of this analysis is to build an optimally-controlled HS which embeds the largest possible quantity of magnetic material. The measurements in Sec.\,\ref{sec:era} and in the first part of Sec.\,\ref{sec:erb} are obtained at room temperature, while the test of the haloscope transducer was performed at $T=90\,$mK. At this temperature, quantum effects prevail over thermal ones, as $k_B T < \hbar \omega$, where $k_B$ is the Boltzman constant.


	\subsection{Single sphere coupling}
	\label{sec:era}
	
In the first study, single spheres of diameters ranging from 0.5 mm to 2.5 mm have been mounted in two different cavities. The TM110 mode frequency is 14.09 GHz for the first cavity and 10.65 GHz for the second, with wavelengths of the rf radiation of 21\,mm and 28\,mm, respectively. Such values allow us to test the system in a regime where the diameter-to-wavelength ratio is larger than 1/10, and dimensional effects should come into play\cite{PhysRev.105.390}.

\begin{figure}
	\includegraphics[width=.5\textwidth]{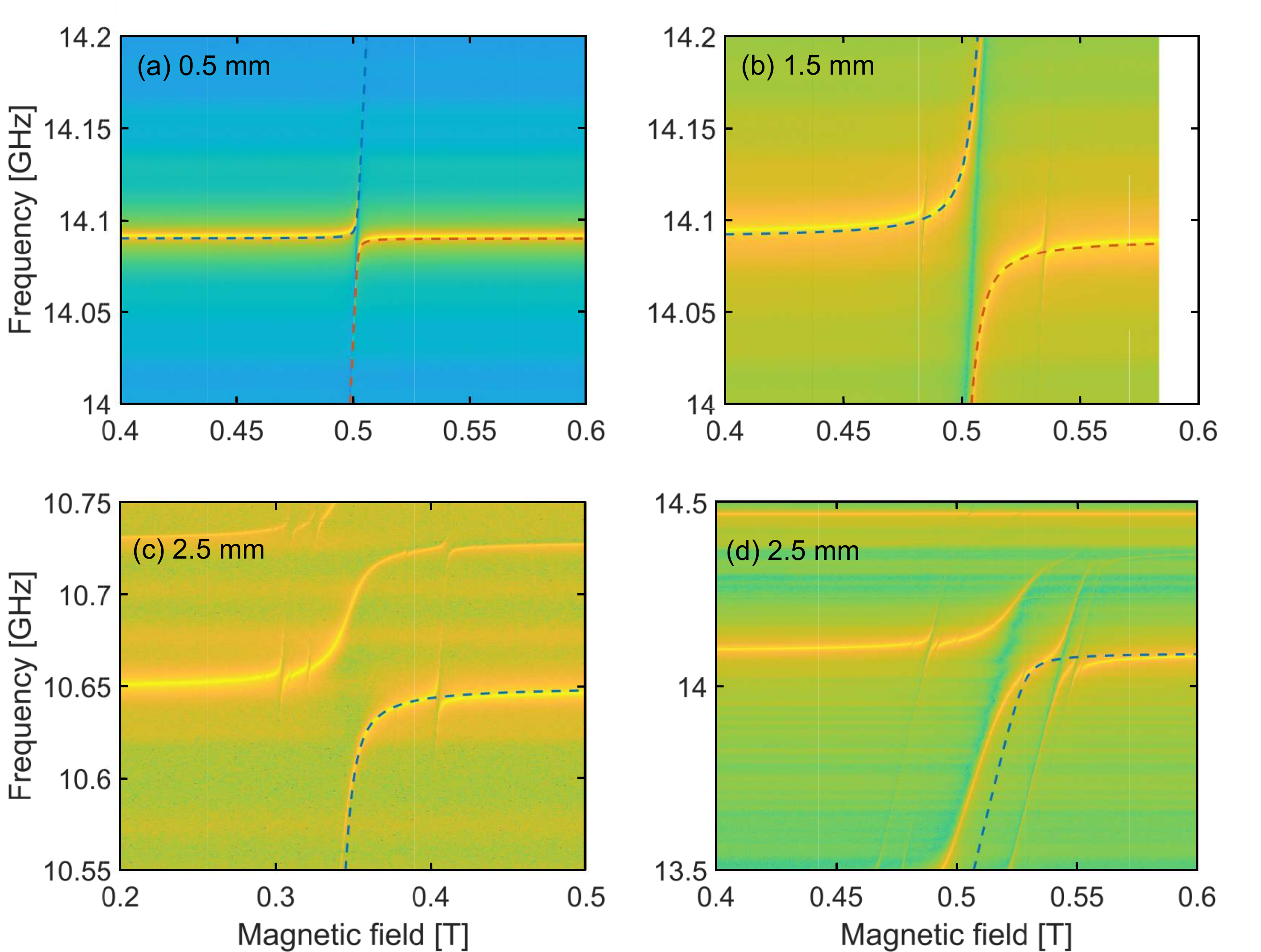}
	\includegraphics[width=.45\textwidth]{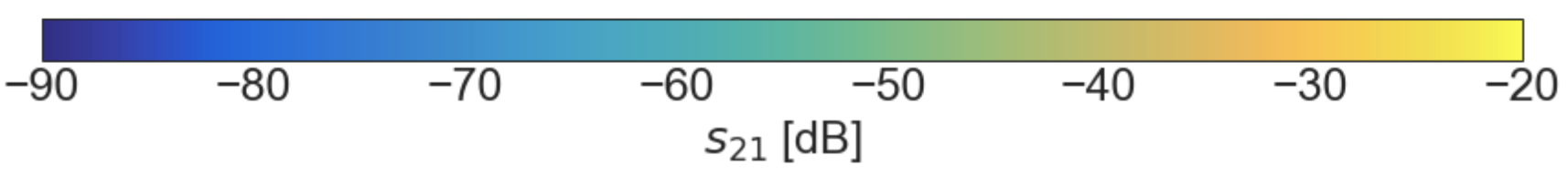}
	\caption{Anticrossing curve of spheres with different diameter. (a) and (b) show the dispersion relation of single spheres of different diameters coupled to the same cavity mode, the coupling strength is obtained by the fit (dashed lines), and scales as expected. (c) and (d) are the anticrossings of the same sphere coupled to the TM110 mode of two different cavities, and the dashed line is the lower frequency hybrid mode. In plot (d) one can see a discrepancy between the theory and the result, possibly due to the fact that the sphere diameter is larger than one tenth of the microwave field wavelength.}
	\label{fig:gslope}
	\end{figure}

Fig.\,\ref{fig:gslope} shows some results in the form of anticrossing curves. We have verified that the hybridization scales as the square root of sample volume, confirming that all the spins are behaving coherently, in agreement with the Tavis-Cummings model. The same behavior is expected when the number of spheres is increased.

One can note that the dispersion relation of Fig.\,\ref{fig:gslope}(c) closely resembles the one of Fig.\,\ref{fig:cdmn}(b), as the coupling strength between the 2.5\,mm sphere and the 10.65\,GHz mode of the cavity is large enough to involve both the $c$ and $d$ polarizations of this TM110 mode. 
As expected from the symmetry of the system, the coupling strengths $g_{cm}$ and $g_{dm}$ are about the same.
In Fig.\,\ref{fig:cdmn}(b) the coupling to the $d$ mode is much weaker because of the characteristic of the antennas, discussed in Section \ref{sec:sqm}.

A first dimensional effect is observed when the largest diameter YIG sphere is coupled to the higher frequency cavity, i.e. the one with smaller wavelength. It consists in an apparent variation of the $g$-factor, as was first evidenced in [\onlinecite{PhysRev.81.477}] for various ferrites.
The 2.5\,mm sphere's electrons, placed in the 14\,GHz cavity, exhibit a gyromagnetic ratio of 26\,GHz/T and not of 28\,GHz/T, as reported in Fig.\,\ref{fig:gslope}(d), which corresponds to an apparent $g$-factor of 1.86.
In the figure, the dashed line shows the expected behavior of the hybrid mode's resonant frequency for $g=2$, which significantly differs from the measured spectra.
The effect is not present if the same sphere is coupled to the lower frequency cavity, as shown in Fig.\,\ref{fig:gslope}(c), were the dashed line is superimposed to the measured hybrid frequency.
The variation of the $g$-factor was observed only when the sample size exceeds one tenth of the radiation field wavelength, in agreement with theoretical suggestions\cite{PhysRev.105.390}.
	
\begin{figure}
	\centering
	\includegraphics[width=.45\textwidth]{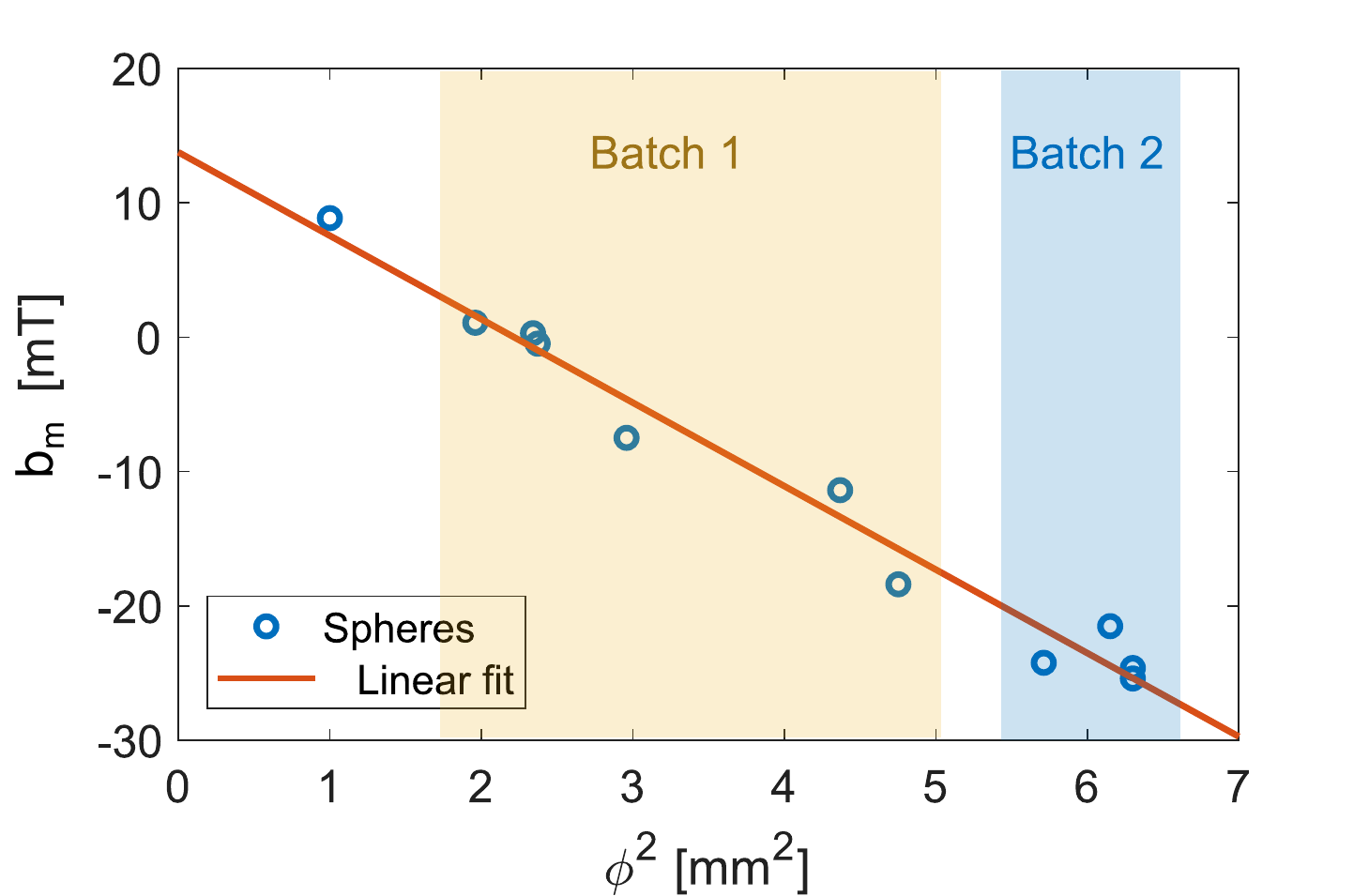}
	\caption{Relation between the sphere's diameter and the full hybridization field $B_{\rm fh}$, measured with the 14\,GHz cavity. The positive offset follows from the zero-field splitting of the YIG.}
	\label{fig:diamB0}
	\end{figure}

A second size effect has also been evidenced in our measurements, and it relates the dimension of the ferrite to its resonant frequency. Such effect was first observed in the fifties\cite{PhysRev.81.477} and described ten years later\cite{PhysRev.81.477,doi:10.1139/p58-114,Krupka_2018}, and states that larger spheres need higher magnetic fields to reach the same resonance frequency.
Here, $B_{\rm fh}$ is the applied magnetic field value that realizes full hybridization, measured through the current flowing in the magnet.
In perfectly spherical samples, the related theory\cite{PhysRev.81.477,doi:10.1139/p58-114,Krupka_2018} introduces a diameter dependent offset frequency
	\begin{equation}
 	b_m(\phi) =  \omega_c/\gamma - B_{\rm fh} = -\frac{2\pi^2  M_0}{45\lambda_m^2}(5+\epsilon_r)\phi^2 + b_0
	\label{eq:deltaomega}
	\end{equation}
where $M_0\simeq178\,$mT and $\epsilon_r$ are the YIG saturation magnetization and relative dielectric constant, $\lambda_m$ is the wavelength associated to the Larmor frequency $\omega_m$,  $\phi$ is the sphere's diameter, and $b_0$ is the zero-diameter offset.
Fig.\,\ref{fig:diamB0} shows the measured values for our set of spheres, together with a linear fit of the data. Using Eq.\,(\ref{eq:deltaomega}), the magnetic field offset for null sphere diameter is found to be $b_0 = 14\pm9$\,mT, compatible with the zero-field splitting calculations performed for YIG\cite{PhysRevB.48.16407}. 
The resulting slope is $6.2\,\mathrm{mT/mm}^2$, 
from which we extract a value of $\epsilon_r\simeq30$, which differs by a factor about 2 from the one found in the literature $\epsilon_r\simeq15$\cite{Krupka_1999,1325655}.
It is true that other measurements of the YIG permittivity yielded values higher\cite{doi:10.1063/1.351846} or lower\cite{doi:10.1063/1.3626057} than 15, however this discrepancy could be explained by some other diameter-dependent size-effect, for which indications may already be present in the X-band measurements of a previous work\cite{PhysRev.81.477}.
Furthermore, as we do not have a precise control over the sample shape, it is not possible to exclude that our spheres are affected by a size-dependent ellipticity that reduces the goodness of Eq.\,(\ref{eq:deltaomega}).

Whilst the second size-effect is not expected to influence the magnon-to-photon transduction, it is not clear whether the variation of the gyromagnetic ratio does.
In the forthcoming measurements, to avoid this issue, we conservatively choose to engineer our HS-based transducer employing spheres almost with the same diameter, namely $\phi\simeq2.1\,$mm.


	\subsection{Multiple spheres acting as one}
	\label{sec:erb}
To move forward in the realization of the transducer, one needs to grasp the coupling of two spheres before scaling up the system to an arbitrarily large number of samples. We focus on understanding the system dispersion relation in the light of the interactions shown in Fig.\,\ref{fig:cdmn}, to maintain the optimal control of the HS, and demonstrate its applicability as an axion to electromagnetic transducer.

First of all, as it is clear from Fig.\,\ref{fig:diamB0}, two spheres with different diameters will exhibit different Larmor frequencies even when subjected to the same magnetic field. An HS with two such spheres would have a dispersion relation as the one described in Fig.\,\ref{fig:cdmn}(d), i.e. equivalent to the situation of two samples with different zero field splitting and with the presence of a dark mode. It follows that, to preserve the control of the HS and the coherent coupling of a large number of spheres, the samples' diameter should be as akin as possible.

\begin{figure}
	\centering
	\includegraphics[width=.45\textwidth]{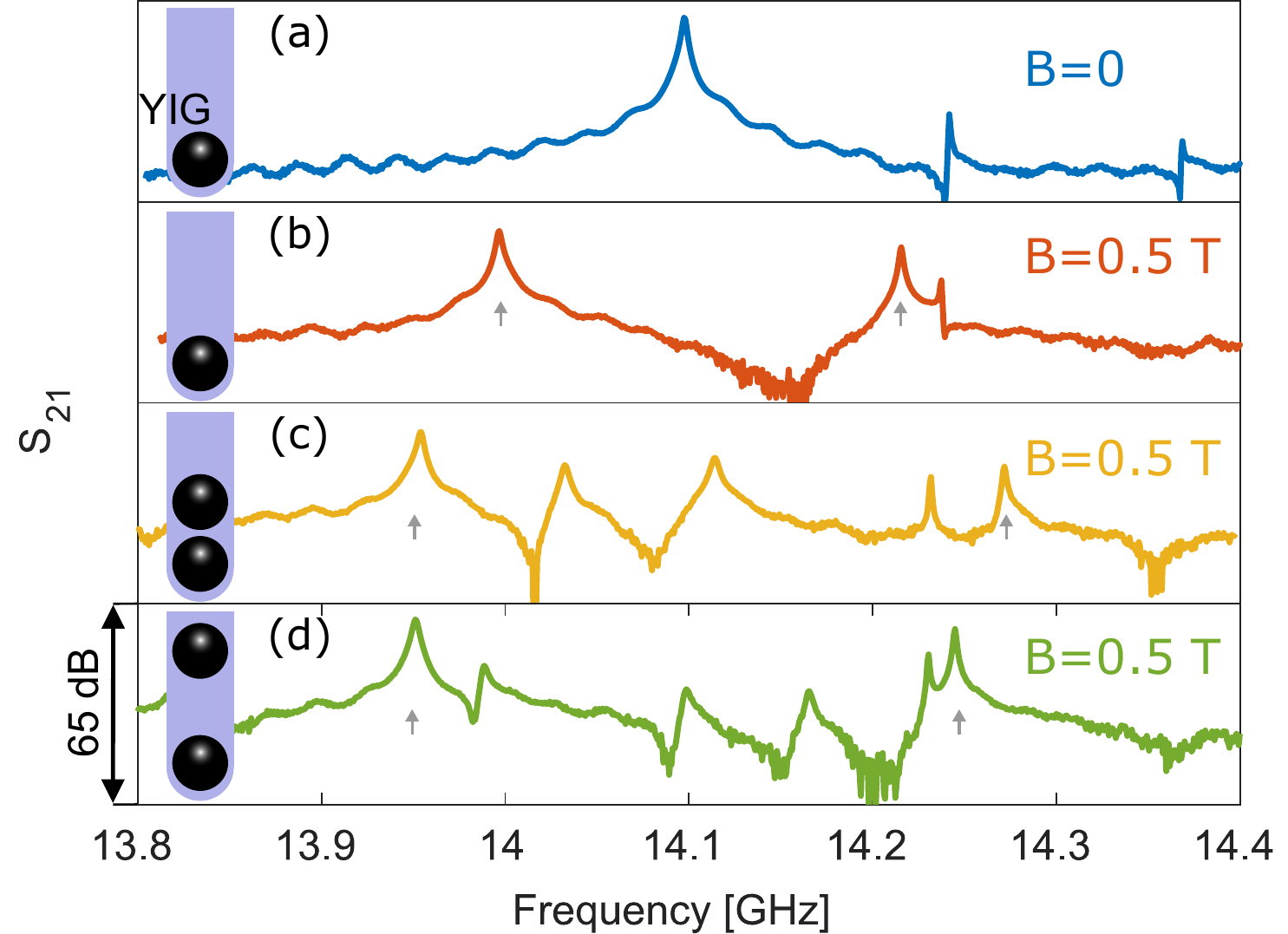}
	\caption{Coherent coupling of two spheres, and effect of their relative distance. On the left side of the plots four schematic drawings show the YIGs (black sphere) position inside the pipe (blue). The pipe is placed on the cavity axis, parallel to the static magnetic field. Plots (a) and (b) show the single sphere configuration without and with magnetic field, that produces the hybrid modes indicated by the small vertical arrows. In (c) one sphere is placed close to the first one, with the effect of increasing the coupling, i.e. the hybrid mode separation, but also of generating a number of dark modes in the intermediate frequency interval. It is evident from (d) that the strength of such modes is reduced by tens of dB thanks to the distance between the two YIGs, in agreement with what is expected from the model.}
	\label{fig:gmn}
	\end{figure}

Secondly, the chance of a magnon bouncing through magnetic modes before being converted to photons has to be minimum, otherwise it will result in a non-negligible signal loss. Magnons in different samples are assumed to interact with each other through a dipole-like potential, hence we studied the sphere-sphere interaction by varying the distance between two YIGs as shown in Fig.\,\ref{fig:gmn},
which can be compared with the theoretical results of Fig.\,\ref{fig:cdmn}(d).
	
The spheres are placed along the cavity axis in both rf and dc uniform magnetic fields.
Being the dc field aligned with the cavity axis, when this is turned on all the spheres will have their easy axis along the same line, i.e. the cavity axis again.
Along this direction, the transmission spectra are independent of the sphere position provided that this is kept in a central range of about 5\,cm length.

In our measurements, we will not care about the absolute positions of the spheres, albeit in their relative separation, which, as we will see, is strongly influencing the results.
Following the plots of Fig.\,\ref{fig:gmn}, in (a) and (b) we obtain the usual hybridization with one sphere by turning on the static field.
Then, in (c), a second sphere is introduced in the system with less than 0.5\,mm of spacing from the first one: the vacuum Rabi splitting increases and extra peaks appear between the two hybrid modes. By increasing the separation between the spheres to $4\,\mathrm{mm}\sim2\phi$, we obtain the plot (d) showing a reduced amplitude of the peaks which are not hybrid modes and a slightly smaller Rabi splitting, now closer to $\sim\sqrt{2}$ times the single sphere case. Relying on the model, described essentially by Figure \ref{fig:cdmn}(d), the plots \,\ref{fig:gmn}(c) and (d) are evidencing the effects of a magnon-magnon coupling for close spheres: an increase of the splitting due to dispersive shifts of the hybrid resonances and a high transmission through the dark modes.
These observations bolster the assumption that the sphere-sphere coupling is dipole-like, as it depends on their distance and, specifically, it is much weakened when the samples are separated by more than their diameter.
	\begin{figure*}
	\centering
	\includegraphics[width=.375\textwidth, valign=c]{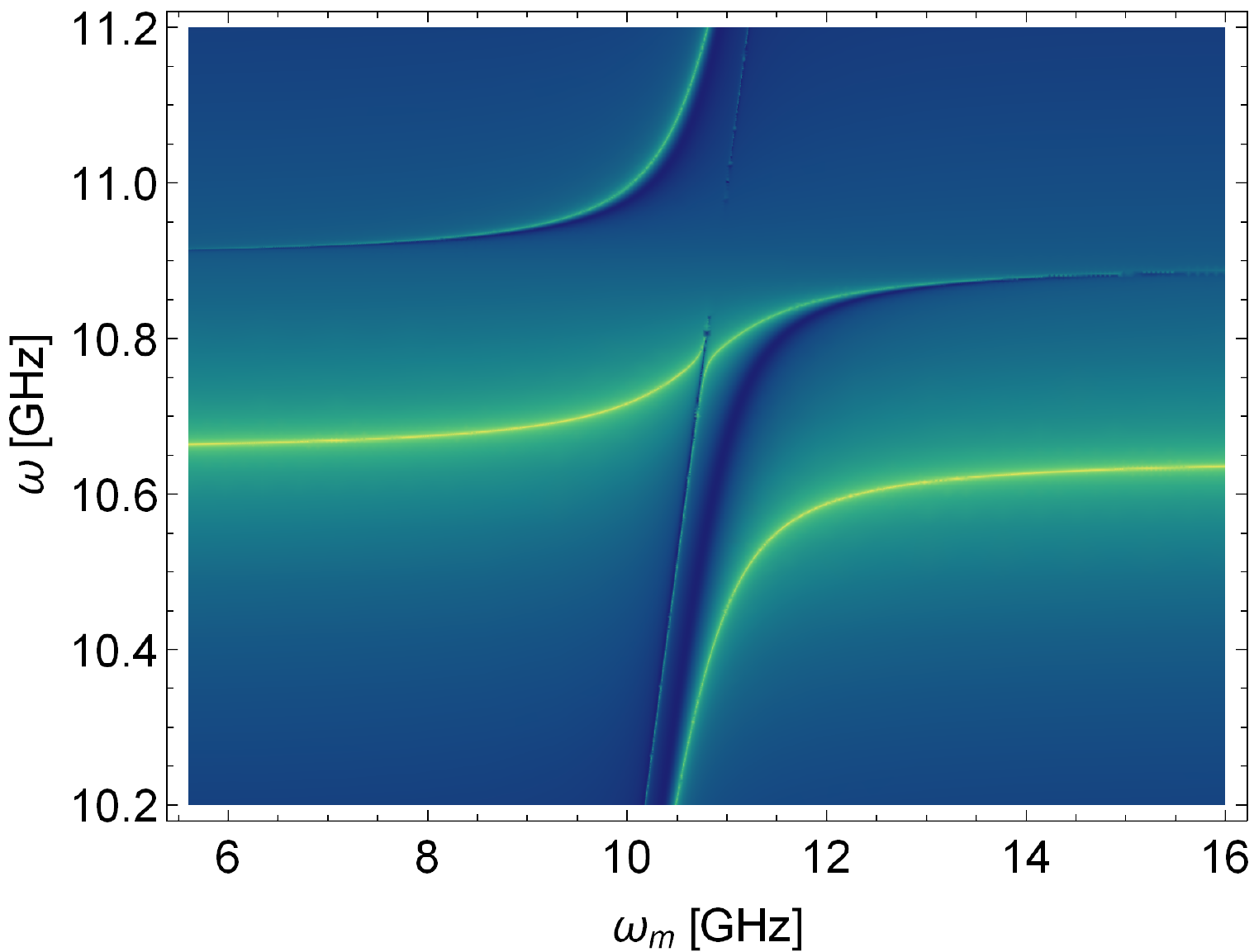}
	\includegraphics[width=.07\textwidth, valign=c]{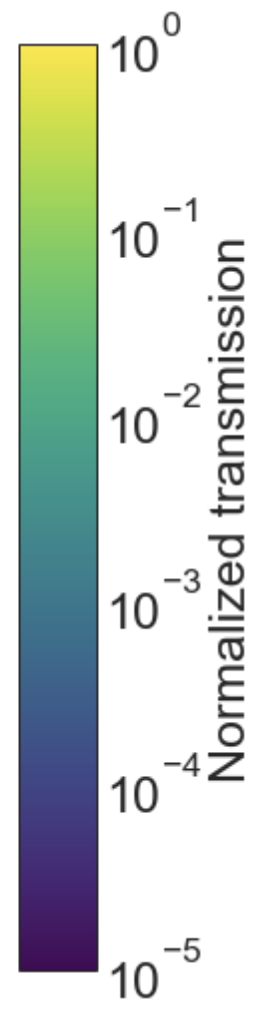}
	\hspace{.5cm}
	\includegraphics[width=.4\textwidth, valign=c]{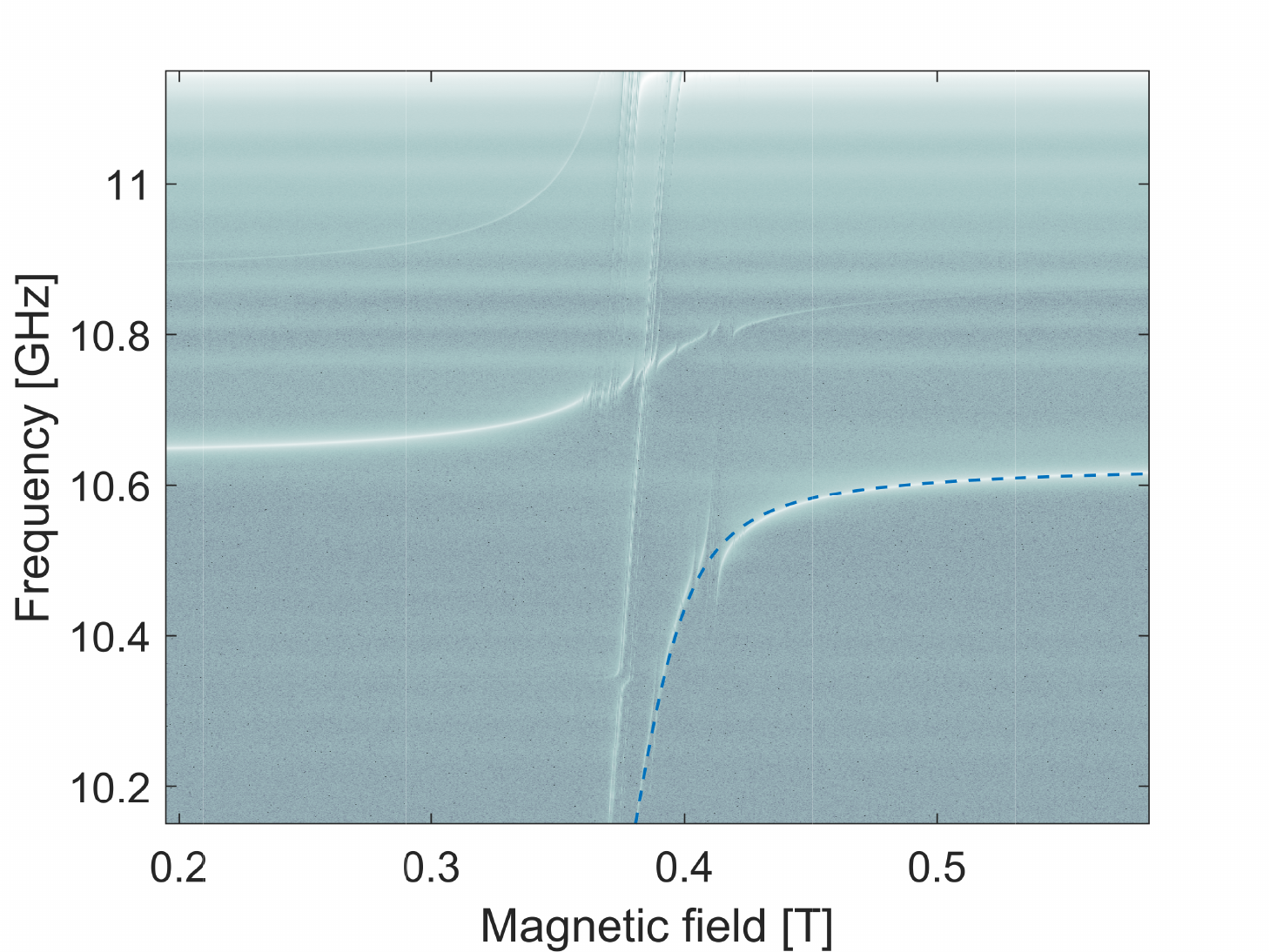}
	\hspace{-0.75cm}
	\includegraphics[width=.07\textwidth, valign=c]{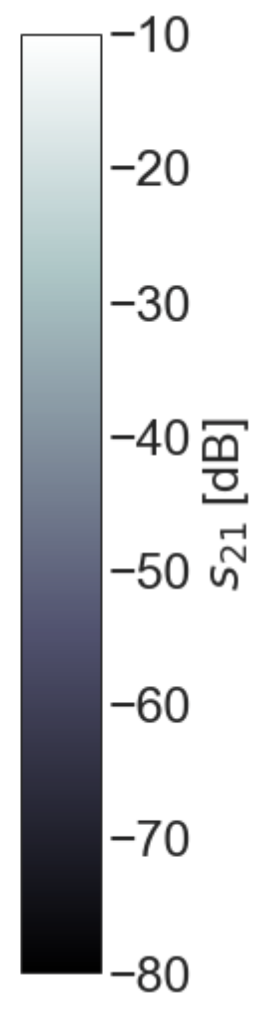}
	\caption{Ten spheres HS: simulated (left, where blue-to-light green corresponds to low-to-high transmission) and measured (right, bright colors correspond to higher transmission) anticrossing dispersion relation, obtained with the model based on ${\cal H}_4$ and a room temperature transmission measurement, respectively. The parameters used to obtained the left anticrossing curve $\omega_-$ are detailed in the text.}
	\label{fig:HMtuning}
	\end{figure*}

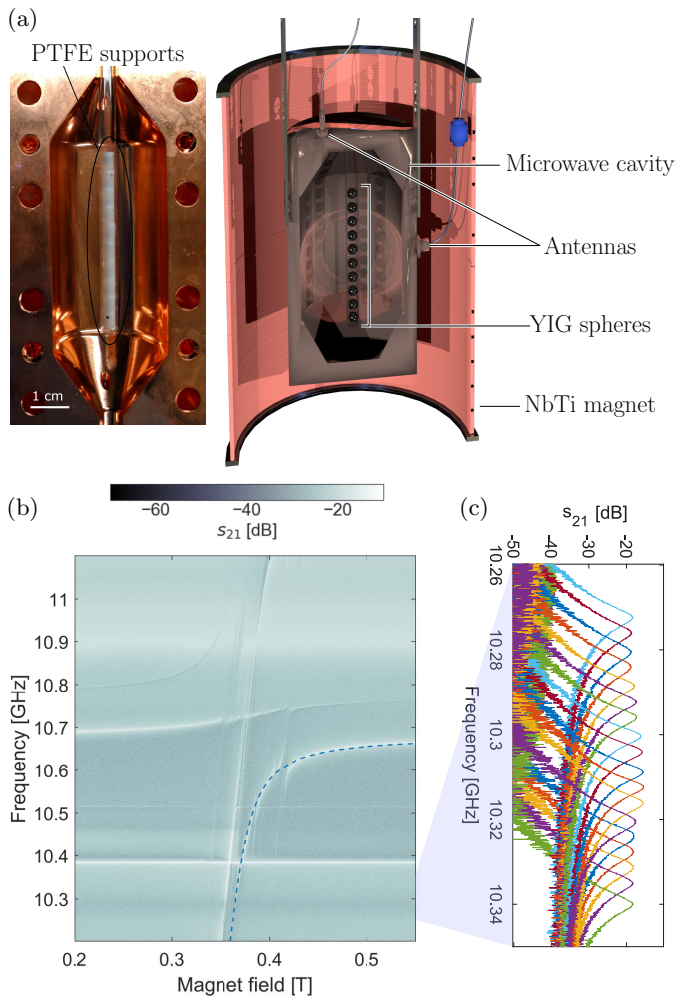
\begin{figure}
	\begin{tikzpicture}
	\node at(0,0){
	\includegraphics[width=.5\textwidth]{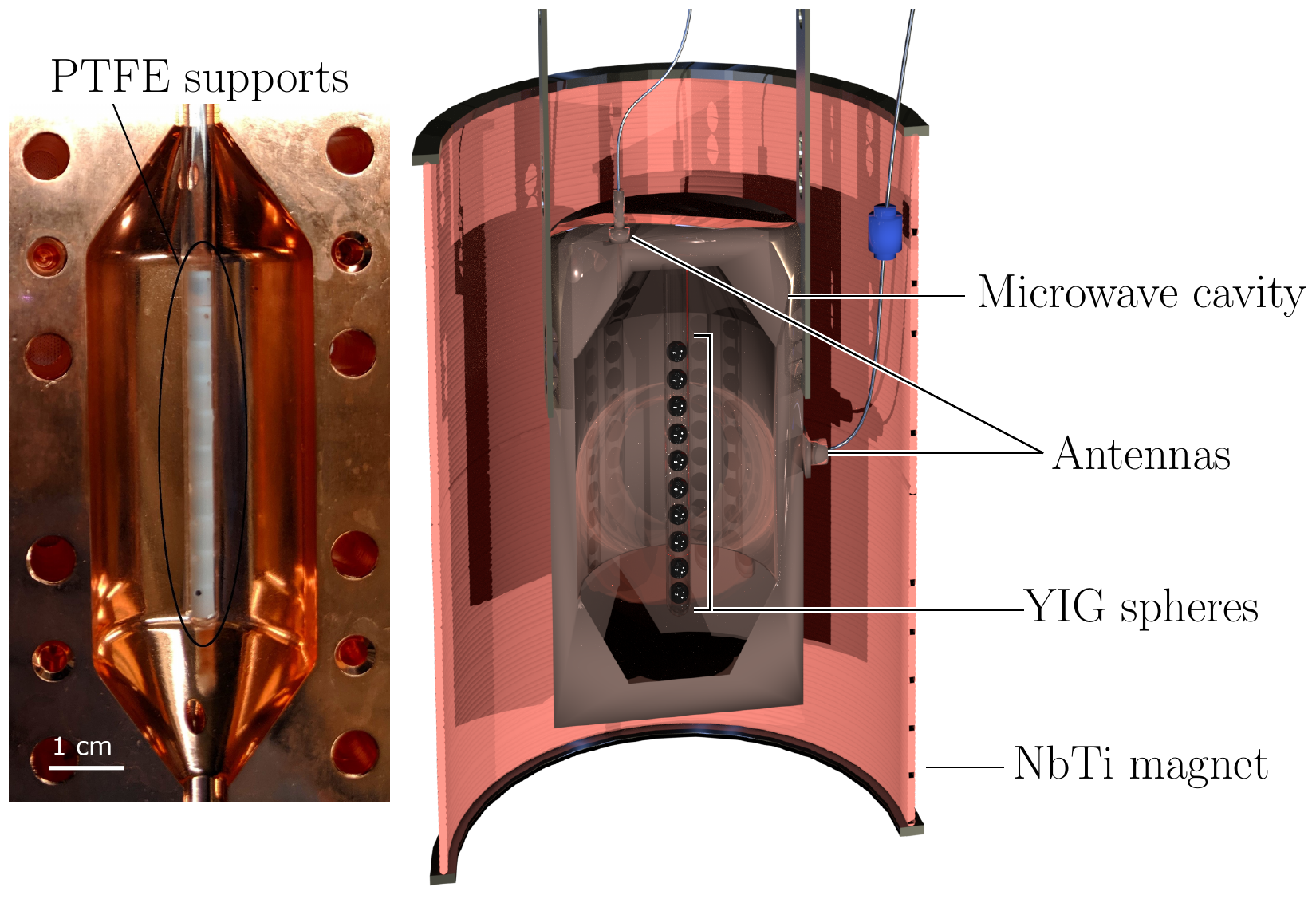}
	};
	\node at(0,-6.5){
	\includegraphics[width=.5\textwidth]{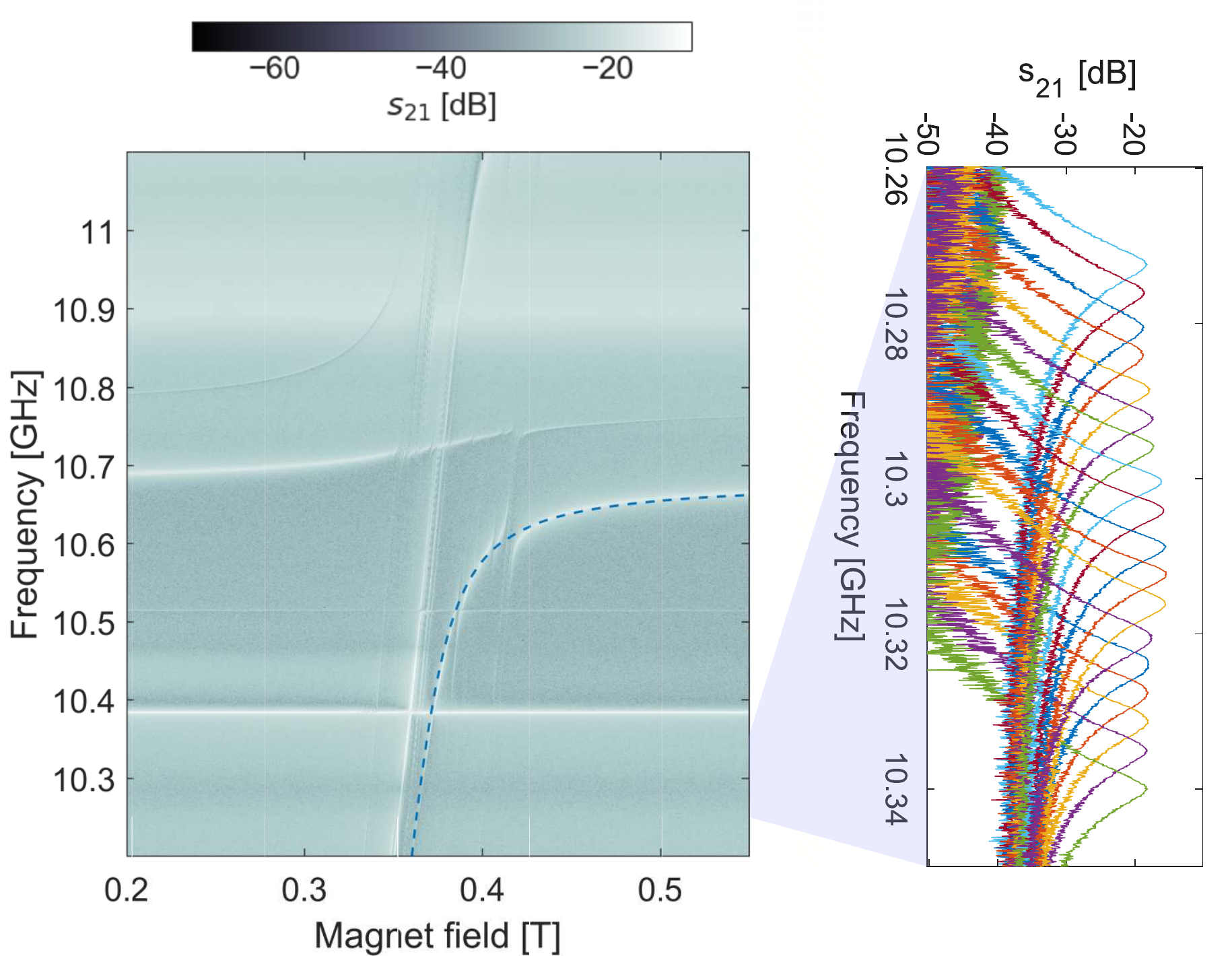}
	};
	\node at(-4.25,3) {(a)};
	\node at(-4.25,-3.5) {(b)};
	\node at(1.75,-3.5) {(c)};
	\end{tikzpicture}
	\caption{(a) Picture (left) and scheme (right) of the HS tested at a temperature of 90\,mK and used in a FH\cite{PhysRevLett.124.171801}. See text and Appendix \ref{app:apparatus} for a detailed description of the apparatus. (b) Anticrossing curve obtained at 90\,mK with the $\omega_-$ frequency evidenced with a dashed line, and (c) tuning of the hybrid mode frequency $\omega_-$ with a field variation of 3\,mT.}
	\label{fig:HSrender}
	\end{figure}

The scaling of the system to a larger number of spheres, necessary to realize a more sensitive ferromagnetic haloscope, is straightforward once
the behavior of the two-YIGs HS is understood.
The conditions considered to engineer such an enhanced HS are summarized as follows:
(i) the single sphere diameter is smaller than 1/10 of $\lambda_m$; (ii) the spheres are free to rotate and align with the external field; (iii) each bare sphere is far in the strong coupling regime with the cavity mode, and is positioned on the cavity axis; (iv) all the spheres must have the same diameter; (v) the relative distance between the spheres should be about $2\phi$.
Following these constraints we filled the volume having uniform magnetic fields, rf of the cavity and dc of the magnet, with ten 2.1 mm diameter YIG spheres. Each one placed along a cylindrical fused silica pipe and separated by thin PTFE caps (see Fig.\,\ref{fig:HSrender} for a sketch and a picture of the system).

We now note that the experimental errors of the measurements are mostly systematic, and possibly due to variations of the samples positioning in the different tests.
As a consequence, the uncertainties in determining the eigenfrequencies of the HS are not related to the frequency resolution, and we estimate them by confronting repeated measurements of the same HS configuration, which are found to be consistent within an 8\% error.

The dispersion relations of Fig.\,\ref{fig:HMtuning}, comparing the theory to the experiment, demonstrate that the resulting HS can be completely described by means of our model based on the Hamiltonian ${\cal H}_4$ in Eq.\,(\ref{eq:Hcdmn}).
The theoretical spectrum is calculated considering only an external coupling to the cavity mode $c$ at 10.65\,GHz (see Section \ref{sec:sqm}).
The experimental spectrum in Fig. \ref{fig:HMtuning} (right) was collected with the use of a movable antenna allowing for a good coupling with mode $c$, and, for symmetry reasons, a much weaker one with mode $d$. Our model yields a sound agreement with the measurements for what concerns mode frequencies and their relative transmission amplitude. For example, the measured transmission of the $d$ mode is extremely low, and the one of the dark mode almost vanishes at higher frequencies.
In the experimental plot, a fit of the lower frequency hybrid mode $\omega_-$ is also shown as a dashed line.
The analysis of the anticrossing curve shows a Rabi splitting of 528\,MHz, correctly scaling as the square root of the number of spheres from the single sphere value of about 180 MHz. 
The resonant frequencies of the cavity modes are the ones reported in Section \ref{sec:sqm}, and remaining parameters are $g_{dm}=264\,$MHz and $g_{mn}=28\,$MHz. The local discrepancies between the two plots (i.\,e. the anticrossing of $\omega_-$ around 0.4\,T) can be imputed to spurious magnon modes.
This mode preserves a good antenna output, and is less affected by the presence of other resonances. In fact, the closer cavity modes are the TM11$z$, for $z=1,2,\dots$, which lie at frequencies higher than $\omega_c$, making $\omega_-$ well isolated.
This mode originates from the coherent coupling of all the ten YIG spheres, and has an adequate coupling to the dipole antenna used to extract the signal, and hence its signal transduction is efficient.

This system, with minor modifications for thermalization (see Appendix \ref{app:apparatus}), was then operated at ultra cryogenic temperatures. The anticrossing pattern obtained at 90\,mK is reported in Fig.\,\ref{fig:HSrender}(b) and closely resembles the one of Fig.\,\ref{fig:HMtuning}, with the increase of the coupling $g_{cm}$ as expected from the higher value of the YIG saturation magnetization at such temperatures. The single sphere splitting increases by about 17\%\cite{PhysRev.134.A1581} and becomes 210\,MHz. A fit with the model based on ${\cal H}_4$ on the experimental dispersion relation gives $2g_{cm} = (2\pi)638$\,MHz, to be compared with the value $210\sqrt{10}\,\mathrm{MHz}=664\,\mathrm{MHz}$. Within the experimental uncertainties, the ten YIG spheres are coherently coupled to the cavity mode, as they effectively react as a single oscillator. Furthermore, the HS behaves as expected also at temperatures where quantum effects dominate over thermal fluctuations\cite{PhysRevLett.113.083603}.

Some other traits of the HS dynamics can be seized by looking at the dispersion relations in Fig.\,\ref{fig:HMtuning} and Fig.\,\ref{fig:HSrender}(b): the coupling of the magnetic $m$-mode with the $d$-mode of the cavity, and the existence of a dark mode related to a magnetic mode $n$. Evidence for these facts comes by looking at the two models described in Fig.\,\ref{fig:cdmn}(b) and Fig.\,\ref{fig:cdmn}(d), respectively.
The explanation for the dark mode is not straightforward.
In addition to the Kittel mode of uniform precession, a ferromagnetic sphere has several higher order magnetostatic modes. Following the mode classification of Walker \cite{doi:10.1063/1.1723117}, we identify the uniform mode $m= (1, 1, 0)$, which is degenerate with the $(4, 3, 0)$ one. The latter could be the $n$-mode producing the dark mode, since sphericity is not perfect in our home made samples, and $m$ and $n$ could actually be slightly detuned from each other\cite{doi:10.1063/1.1723117}. However, we are actually not favoring such explanations since single spheres dispersion curves only show standard behavior.
The presence of a mode $n$ could also be due to the use of spheres with different diameters, as it was discussed in Section \ref{sec:er} (see Fig.\,\ref{fig:diamB0}). Also for this case, we believe this is not the correct explanation. In fact, by performing measurements with two spheres we have verified that, for our set, coherent interaction with all the available spins is not realized only when the differences in sphere diameters satisfy $\gamma b_m(\phi)\gg\gamma_m$ (see discussion of Fig. \ref{fig:cdmn}(d) and in Section \ref{sec:sqm}). Since all the YIGs used for the ultra cryogenic measurement have much close diameters, we believe that the dark mode in our anticrossing curves of Fig.\,\ref{fig:HMtuning} and \ref{fig:HSrender}(b) results from some residual magnon-magnon coupling. A way to certify this could be testing the presence of the dark mode with almost identical spheres.
In Fig.\,\ref{fig:HSrender}(b) are present two low frequency modes which are absent in the room temperature anticrossing of Fig.\,\ref{fig:HMtuning}. This unidentified cavity resonances are visible also at 300\,K, but their frequency shifts with the temperature. Even if they appear in the spectra, they are not coupled to magnon modes and so does not interfere with the FH operation.

As the apparatus behavior is consistently described by the model based on ${\cal H}_4$, we can now use it to find the maximal transduction bandwidth of the realized set-up for an input on the magnetic mode $m$ and readout on the cavity mode $c$ with matched antenna. This is the general situation of a ferrimagnetic haloscope described in the introduction, where a pseudo-magnetic field due to a dark matter axion excites a magnetic resonance in the material, producing an rf power on the coupled cavity mode. By tuning the Larmor frequency $\omega_m$ through the magnetic field it is possible to search for dark matter fields oscillating at various frequencies $\omega_-$, considering a transduction efficiency given by the relative change of magnon-to-photon conversion probability. We can define the detection bandwidth as the range of frequencies for which transduction efficiency is within 3 dB from the maximum; we mention that the verification of this approach validity will be presented in a forthcoming work\cite{inprep}. For the apparatus described by Fig.\,\ref{fig:HSrender}(b) this bandwidth is about 400 MHz. In Fig.\,\ref{fig:HSrender}(c) a close up of the tuning obtained with a variation of 3 mT of the magnetic field is shown. The possibility to achieve large and simple tunability for the input frequency is a major advantage of a ferromagnetic haloscope\cite{PhysRevLett.124.171801}, and can be used to simplify many HS-based applications\cite{Clerk2020,yigbell}.


	\section{Further developments and conclusions}
	\label{sec:fdc}
Detection sensitivity of FH is directly related to the amount of sensitive material. To improve sensitivity, while keeping the same central working frequency, longer cylindrical cavities holding more spheres can be used, and off axis loading of the spheres can be implemented. In order to coherently use all the available spins, each sphere must be strongly coupled to the cavity mode\cite{inprep}. This results in the production of ultra strong vacuum Rabi splitting for the complete system, with the drawback of possible interference with other cavity modes or between higher order magnetic modes. To avoid such problems, we leverage the fact that as long as the single sphere coupling is larger than the losses the transduction efficiency is preserved.
Hence, magnetic spheres can be placed in a region of lower rf magnetic field of the cavity mode, thereby reducing the single sphere coupling to a few times the magnon or cavity linewidth, with the result of a smaller total splitting.
Moreover, some preliminary measurements that we have performed show that magnon-magnon interaction is kept small when two or more spheres are placed in a plane perpendicular to the cavity axis, thus allowing a small gap separation between the spheres in such direction. It is then possible to foresee a single microwave cavity with a volume of active material one order of magnitude larger than the one employed in this article, just by filling the cavity with planes of spheres, each plane separated by the required distance to avoid mutual interaction.
If carefully devised, these future HSs may still be described by a few oscillators models, preserving their optimal control, and remarkably increasing their possible applications.
	

In conclusion, we outlined a model composed of an arbitrary number of photon and magnon oscillators.
The analysis was limited to the experimental case of two cavity modes and two magnetic modes, and the various couplings among them were separately studied.
Afterwards, we make use of a flexible HS to experimentally reproduce the theoretical results and understand the constraints to take into account when devising a HS with a considerable material quantity. The outcome is a recipe that we use to build the largest optimally-controlled HS to date, which we test at room and at millikelvin temperature.
This device is used as axion-to-electromagnetic field transducer in a FH\cite{PhysRevLett.124.171801}, paving the way to the use of these devices to search for Dark Matter with a cosmologically relevant sensitivity. Our results benchmark a starting point for future large HS designs, and demonstrates that the degrees of freedom of such a complex system can be reduced to an acceptable number.

\begin{acknowledgements}
The authors would like to deeply acknowledge Enrico Berto for the help in the fabrication of the spheres, and Riccardo Barbieri for the initial formulation of the model. We are thankful to Andrea Benato, Fulvio Calaon and Mario Tessaro for the work on the cryogenics and electronics of the setup, and to David Alesini and Mario Zago for the cavity design and realization. We eventually acknowledge Laboratori Nazionali di Legnaro for hosting the experimental apparatus and for providing us with liquid helium on demand.
\end{acknowledgements}

\bibliographystyle{apssamp}
\bibliography{nCoherentSpheres}

\begin{thebibliography}{10}

\bibitem{haroche}
Serge Haroche and Jean-Michel Raimond.
\newblock {\em Exploring the Quantum: Atoms, Cavities, and Photons}.
\newblock Oxford University Press, 2006.

\bibitem{Clerk2020}
A.~A. Clerk, K.~W. Lehnert, P.~Bertet, J.~R. Petta, and Y.~Nakamura.
\newblock Hybrid quantum systems with circuit quantum electrodynamics.
\newblock {\em Nature Physics}, 16(3):257--267, Mar 2020.

\bibitem{Lachance_Quirion_2019}
Dany Lachance-Quirion, Yutaka Tabuchi, Arnaud Gloppe, Koji Usami, and Yasunobu
  Nakamura.
\newblock Hybrid quantum systems based on magnonics.
\newblock {\em Applied Physics Express}, 12(7):070101, jun 2019.

\bibitem{Kittel2004}
Charles Kittel.
\newblock {\em Introduction to Solid State Physics}.
\newblock Wiley, 8 edition, 2004.

\bibitem{PhysRev.143.372}
Y.~R. Shen and N.~Bloembergen.
\newblock Interaction between light waves and spin waves.
\newblock {\em Phys. Rev.}, 143:372--384, Mar 1966.

\bibitem{PhysRevLett.111.127003}
Hans Huebl, Christoph~W. Zollitsch, Johannes Lotze, Fredrik Hocke, Moritz
  Greifenstein, Achim Marx, Rudolf Gross, and Sebastian T.~B. Goennenwein.
\newblock High cooperativity in coupled microwave resonator ferrimagnetic
  insulator hybrids.
\newblock {\em Phys. Rev. Lett.}, 111:127003, Sep 2013.

\bibitem{PhysRevLett.113.083603}
Yutaka Tabuchi, Seiichiro Ishino, Toyofumi Ishikawa, Rekishu Yamazaki, Koji
  Usami, and Yasunobu Nakamura.
\newblock Hybridizing ferromagnetic magnons and microwave photons in the
  quantum limit.
\newblock {\em Phys. Rev. Lett.}, 113:083603, Aug 2014.

\bibitem{PhysRevApplied.2.054002}
Maxim Goryachev, Warrick~G. Farr, Daniel~L. Creedon, Yaohui Fan, Mikhail
  Kostylev, and Michael~E. Tobar.
\newblock High-cooperativity cavity qed with magnons at microwave frequencies.
\newblock {\em Phys. Rev. Applied}, 2:054002, Nov 2014.

\bibitem{Zhang2015}
Dengke Zhang, Xin-Ming Wang, Tie-Fu Li, Xiao-Qing Luo, Weidong Wu, Franco Nori,
  and J.~Q. You.
\newblock Cavity quantum electrodynamics with ferromagnetic magnons in a small
  yttrium-iron-garnet sphere.
\newblock {\em npj Quantum Information}, 1(1):15014, 2015.

\bibitem{Ladd2010}
T.~D. Ladd, F.~Jelezko, R.~Laflamme, Y.~Nakamura, C.~Monroe, and J.~L. O'Brien.
\newblock Quantum computers.
\newblock {\em Nature}, 464(7285):45--53, Mar 2010.

\bibitem{Kimble2008}
H.~J. Kimble.
\newblock The quantum internet.
\newblock {\em Nature}, 453(7198):1023--1030, Jun 2008.

\bibitem{RevModPhys.87.1379}
Andreas Reiserer and Gerhard Rempe.
\newblock Cavity-based quantum networks with single atoms and optical photons.
\newblock {\em Rev. Mod. Phys.}, 87:1379--1418, Dec 2015.

\bibitem{RevModPhys.89.035002}
C.~L. Degen, F.~Reinhard, and P.~Cappellaro.
\newblock Quantum sensing.
\newblock {\em Rev. Mod. Phys.}, 89:035002, Jul 2017.

\bibitem{PhysRevLett.113.156401}
Xufeng Zhang, Chang-Ling Zou, Liang Jiang, and Hong~X. Tang.
\newblock Strongly coupled magnons and cavity microwave photons.
\newblock {\em Phys. Rev. Lett.}, 113:156401, Oct 2014.

\bibitem{PhysRevLett.114.226402}
L.~V. Abdurakhimov, Yu.~M. Bunkov, and D.~Konstantinov.
\newblock Normal-mode splitting in the coupled system of hybridized nuclear
  magnons and microwave photons.
\newblock {\em Phys. Rev. Lett.}, 114:226402, Jun 2015.

\bibitem{PhysRevLett.105.140501}
D.~I. Schuster, A.~P. Sears, E.~Ginossar, L.~DiCarlo, L.~Frunzio, J.~J.~L.
  Morton, H.~Wu, G.~A.~D. Briggs, B.~B. Buckley, D.~D. Awschalom, and R.~J.
  Schoelkopf.
\newblock High-cooperativity coupling of electron-spin ensembles to
  superconducting cavities.
\newblock {\em Phys. Rev. Lett.}, 105:140501, Sep 2010.

\bibitem{Tabuchi405}
Yutaka Tabuchi, Seiichiro Ishino, Atsushi Noguchi, Toyofumi Ishikawa, Rekishu
  Yamazaki, Koji Usami, and Yasunobu Nakamura.
\newblock Coherent coupling between a ferromagnetic magnon and a
  superconducting qubit.
\newblock {\em Science}, 349(6246):405--408, 2015.

\bibitem{wang2019quantum}
Yi-Pu Wang, Guo-Qiang Zhang, Da~Xu, Tie-Fu Li, Shi-Yao Zhu, J.~S. Tsai, and
  J.~Q. You.
\newblock Quantum simulation of the fermion-boson composite quasi-particles
  with a driven qubit-magnon hybrid quantum system, arXiv1903.12498.

\bibitem{TABUCHI2016729}
Yutaka Tabuchi, Seiichiro Ishino, Atsushi Noguchi, Toyofumi Ishikawa, Rekishu
  Yamazaki, Koji Usami, and Yasunobu Nakamura.
\newblock Quantum magnonics: The magnon meets the superconducting qubit.
\newblock {\em Comptes Rendus Physique}, 17(7):729 -- 739, 2016.
\newblock Quantum microwaves / Micro-ondes quantiques.

\bibitem{doi:10.1080/09500340.2016.1148212}
Khabat Heshami, Duncan~G. England, Peter~C. Humphreys, Philip~J. Bustard,
  Victor~M. Acosta, Joshua Nunn, and Benjamin~J. Sussman.
\newblock Quantum memories: emerging applications and recent advances.
\newblock {\em Journal of Modern Optics}, 63(20):2005--2028, 2016.

\bibitem{Kurizki3866}
Gershon Kurizki, Patrice Bertet, Yuimaru Kubo, Klaus M{\o}lmer, David
  Petrosyan, Peter Rabl, and J{\"o}rg Schmiedmayer.
\newblock Quantum technologies with hybrid systems.
\newblock {\em Proceedings of the National Academy of Sciences},
  112(13):3866--3873, 2015.

\bibitem{cate}
Caterina Braggio, G~Carugno, Marco Guarise, A~Ortolan, and G~Ruoso.
\newblock Optical manipulation of a magnon-photon hybrid system.
\newblock 118, 09 2016.

\bibitem{PhysRevA.92.062313}
Xavier Fernandez-Gonzalvo, Yu-Hui Chen, Chunming Yin, Sven Rogge, and Jevon~J.
  Longdell.
\newblock Coherent frequency up-conversion of microwaves to the optical
  telecommunications band in an er:yso crystal.
\newblock {\em Phys. Rev. A}, 92:062313, Dec 2015.

\bibitem{PhysRevLett.113.203601}
Lewis~A. Williamson, Yu-Hui Chen, and Jevon~J. Longdell.
\newblock Magneto-optic modulator with unit quantum efficiency.
\newblock {\em Phys. Rev. Lett.}, 113:203601, Nov 2014.

\bibitem{PhysRevB.93.174427}
R.~Hisatomi, A.~Osada, Y.~Tabuchi, T.~Ishikawa, A.~Noguchi, R.~Yamazaki,
  K.~Usami, and Y.~Nakamura.
\newblock Bidirectional conversion between microwave and light via
  ferromagnetic magnons.
\newblock {\em Phys. Rev. B}, 93:174427, May 2016.

\bibitem{Lachance-Quirion425}
Dany Lachance-Quirion, Samuel~Piotr Wolski, Yutaka Tabuchi, Shingo Kono, Koji
  Usami, and Yasunobu Nakamura.
\newblock Entanglement-based single-shot detection of a single magnon with a
  superconducting qubit.
\newblock {\em Science}, 367(6476):425--428, 2020.

\bibitem{Chumak2015}
A.~V. Chumak, V.~I. Vasyuchka, A.~A. Serga, and B.~Hillebrands.
\newblock Magnon spintronics.
\newblock {\em Nature Physics}, 11(6):453--461, Jun 2015.

\bibitem{moiseyev2011non}
Nimrod Moiseyev.
\newblock {\em Non-Hermitian Quantum Mechanics}.
\newblock Cambridge University Press, 2011.

\bibitem{Bender_2007}
Carl~M Bender.
\newblock Making sense of non-hermitian hamiltonians.
\newblock {\em Reports on Progress in Physics}, 70(6):947--1018, may 2007.

\bibitem{Heiss_2012}
W~D Heiss.
\newblock The physics of exceptional points.
\newblock {\em Journal of Physics A: Mathematical and Theoretical},
  45(44):444016, oct 2012.

\bibitem{PhysRevX.6.021007}
Kun Ding, Guancong Ma, Meng Xiao, Z.~Q. Zhang, and C.~T. Chan.
\newblock Emergence, coalescence, and topological properties of multiple
  exceptional points and their experimental realization.
\newblock {\em Phys. Rev. X}, 6:021007, Apr 2016.

\bibitem{Zhang2017}
Dengke Zhang, Xiao-Qing Luo, Yi-Pu Wang, Tie-Fu Li, and J.~Q. You.
\newblock Observation of the exceptional point in cavity magnon-polaritons.
\newblock {\em Nature Communications}, 8(1):1368, Nov 2017.

\bibitem{PhysRevLett.123.237202}
Xufeng Zhang, Kun Ding, Xianjing Zhou, Jing Xu, and Dafei Jin.
\newblock Experimental observation of an exceptional surface in synthetic
  dimensions with magnon polaritons.
\newblock {\em Phys. Rev. Lett.}, 123:237202, Dec 2019.

\bibitem{PhysRevB.99.054404}
Guo-Qiang Zhang and J.~Q. You.
\newblock Higher-order exceptional point in a cavity magnonics system.
\newblock {\em Phys. Rev. B}, 99:054404, Feb 2019.

\bibitem{PhysRevB.99.214415}
Yunshan Cao and Peng Yan.
\newblock Exceptional magnetic sensitivity of
  $\mathcal{P}\mathcal{T}$-symmetric cavity magnon polaritons.
\newblock {\em Phys. Rev. B}, 99:214415, Jun 2019.

\bibitem{PhysRevLett.124.053602}
H.~Y. Yuan, Peng Yan, Shasha Zheng, Q.~Y. He, Ke~Xia, and Man-Hong Yung.
\newblock Steady bell state generation via magnon-photon coupling.
\newblock {\em Phys. Rev. Lett.}, 124:053602, Feb 2020.

\bibitem{PhysRevLett.120.057202}
Yi-Pu Wang, Guo-Qiang Zhang, Dengke Zhang, Tie-Fu Li, C.-M. Hu, and J.~Q. You.
\newblock Bistability of cavity magnon polaritons.
\newblock {\em Phys. Rev. Lett.}, 120:057202, Jan 2018.

\bibitem{PhysRevLett.123.127202}
Yi-Pu Wang, J.~W. Rao, Y.~Yang, Peng-Chao Xu, Y.~S. Gui, B.~M. Yao, J.~Q. You,
  and C.-M. Hu.
\newblock Nonreciprocity and unidirectional invisibility in cavity magnonics.
\newblock {\em Phys. Rev. Lett.}, 123:127202, Sep 2019.

\bibitem{PhysRevA.93.021803}
N.~J. Lambert, J.~A. Haigh, S.~Langenfeld, A.~C. Doherty, and A.~J. Ferguson.
\newblock Cavity-mediated coherent coupling of magnetic moments.
\newblock {\em Phys. Rev. A}, 93:021803, Feb 2016.

\bibitem{yigbell}
Neil Salmon and Stephen Hoon.
\newblock A millimeter-wave bell test using a ferrite parametric amplifier and
  a homodyne interferometer.
\newblock {\em Journal of Magnetism and Magnetic Materials}, 501:166435, 05
  2020.

\bibitem{quaxepjc}
{Crescini, N.}, {Alesini, D.}, {Braggio, C.}, {Carugno, G.}, {Di Gioacchino,
  D.}, {Gallo, C. S.}, {Gambardella, U.}, {Gatti, C.}, {Iannone, G.}, {Lamanna,
  G.}, {Ligi, C.}, {Lombardi, A.}, {Ortolan, A.}, {Pagano, S.}, {Pengo, R.},
  {Ruoso, G.}, {Speake, C. C.}, and {Taffarello, L.}
\newblock Operation of a ferromagnetic axion haloscope at
  $m_a=58\,\mu\mathrm{eV}$.
\newblock {\em Eur. Phys. J. C}, 78(9):703, 2018.

\bibitem{FLOWER2019100306}
Graeme Flower, Jeremy Bourhill, Maxim Goryachev, and Michael~E. Tobar.
\newblock Broadening frequency range of a ferromagnetic axion haloscope with
  strongly coupled cavity–magnon polaritons.
\newblock {\em Physics of the Dark Universe}, 25:100306, 2019.

\bibitem{BARBIERI1989357}
R~Barbieri, M~Cerdonio, G~Fiorentini, and S~Vitale.
\newblock Axion to magnon conversion. a scheme for the detection of galactic
  axions.
\newblock {\em Physics Letters B}, 226(3):357 -- 360, 1989.

\bibitem{PhysRevB.97.014419}
Babak Zare~Rameshti and Gerrit E.~W. Bauer.
\newblock Indirect coupling of magnons by cavity photons.
\newblock {\em Phys. Rev. B}, 97:014419, Jan 2018.

\bibitem{PhysRevB.99.094407}
Y.~Xiao, X.~H. Yan, Y.~Zhang, V.~L. Grigoryan, C.~M. Hu, H.~Guo, and K.~Xia.
\newblock Magnon dark mode of an antiferromagnetic insulator in a microwave
  cavity.
\newblock {\em Phys. Rev. B}, 99:094407, Mar 2019.

\bibitem{PhysRevResearch.1.023021}
Zhedong Zhang, Marlan~O. Scully, and Girish~S. Agarwal.
\newblock Quantum entanglement between two magnon modes via kerr nonlinearity
  driven far from equilibrium.
\newblock {\em Phys. Rev. Research}, 1:023021, Sep 2019.

\bibitem{PhysRevLett.121.087204}
\O{}yvind Johansen and Arne Brataas.
\newblock Nonlocal coupling between antiferromagnets and ferromagnets in
  cavities.
\newblock {\em Phys. Rev. Lett.}, 121:087204, Aug 2018.

\bibitem{walls2007quantum}
Daniel~F Walls and Gerard~J Milburn.
\newblock {\em Quantum optics}.
\newblock Springer Science \& Business Media, 2007.

\bibitem{scully1999quantum}
Marlan~O Scully and M~Suhail Zubairy.
\newblock {\em Quantum optics}.
\newblock American Association of Physics Teachers, 1999.

\bibitem{PhysRevLett.103.083601}
J.~M. Fink, R.~Bianchetti, M.~Baur, M.~G\"oppl, L.~Steffen, S.~Filipp, P.~J.
  Leek, A.~Blais, and A.~Wallraff.
\newblock Dressed collective qubit states and the tavis-cummings model in
  circuit qed.
\newblock {\em Phys. Rev. Lett.}, 103:083601, Aug 2009.

\bibitem{PhysRev.170.379}
Michael Tavis and Frederick~W. Cummings.
\newblock Exact solution for an $n$-molecule-radiation-field hamiltonian.
\newblock {\em Phys. Rev.}, 170:379--384, Jun 1968.

\bibitem{Princep2017}
Andrew~J. Princep, Russell~A. Ewings, Simon Ward, Sandor T{\'o}th, Carsten
  Dubs, Dharmalingam Prabhakaran, and Andrew~T. Boothroyd.
\newblock The full magnon spectrum of yttrium iron garnet.
\newblock {\em npj Quantum Materials}, 2(1):63, Nov 2017.

\bibitem{PhysRevB.101.014439}
Angelo Leo, Anna~Grazia Monteduro, Silvia Rizzato, Luigi Martina, and Giuseppe
  Maruccio.
\newblock Identification and time-resolved study of ferrimagnetic spin-wave
  modes in a microwave cavity in the strong-coupling regime.
\newblock {\em Phys. Rev. B}, 101:014439, Jan 2020.

\bibitem{PhysRev.105.390}
L.~R. Walker.
\newblock Magnetostatic modes in ferromagnetic resonance.
\newblock {\em Phys. Rev.}, 105:390--399, Jan 1957.

\bibitem{PhysRev.81.477}
W.~A. Yager, F.~R. Merritt, and Charles Guillaud.
\newblock Ferromagnetic resonance in various ferrites.
\newblock {\em Phys. Rev.}, 81:477--478, Feb 1951.

\bibitem{doi:10.1063/1.2185878}
J.~E. Mercereau.
\newblock Ferromagnetic resonance g factor to order (kr0)2.
\newblock {\em Journal of Applied Physics}, 30(4):S184--S185, 1959.

\bibitem{doi:10.1139/p58-114}
R.~A. Hurd.
\newblock The magnetic fields of a ferrite ellipsoid.
\newblock {\em Canadian Journal of Physics}, 36(8):1072--1083, 1958.

\bibitem{Krupka_2018}
Jerzy Krupka.
\newblock Measurement of the complex permittivity, initial permeability,
  permeability tensor and ferromagnetic linewidth of gyromagnetic materials.
\newblock {\em Measurement Science and Technology}, 29(9):092001, jul 2018.

\bibitem{doi:10.1063/1.5121618}
M.~X. Bi, X.~H. Yan, Y.~Xiao, and C.~J. Dai.
\newblock Magnon dark mode in a strong driving microwave cavity.
\newblock {\em Journal of Applied Physics}, 126(17):173902, 2019.

\bibitem{PhysRevD.99.101101}
D.~Alesini, C.~Braggio, G.~Carugno, N.~Crescini, D.~D'Agostino,
  D.~Di~Gioacchino, R.~Di~Vora, P.~Falferi, S.~Gallo, U.~Gambardella, C.~Gatti,
  G.~Iannone, G.~Lamanna, C.~Ligi, A.~Lombardi, R.~Mezzena, A.~Ortolan,
  R.~Pengo, N.~Pompeo, A.~Rettaroli, G.~Ruoso, E.~Silva, C.~C. Speake,
  L.~Taffarello, and S.~Tocci.
\newblock Galactic axions search with a superconducting resonant cavity.
\newblock {\em Phys. Rev. D}, 99:101101, May 2019.

\bibitem{CHEREPANOV199381}
The saga of yig: Spectra, thermodynamics, interaction and relaxation of magnons
  in a complex magnet.
\newblock {\em Physics Reports}, 229(3):81 -- 144, 1993.

\bibitem{doi:10.1063/1.1723117}
L.~R. Walker.
\newblock Resonant modes of ferromagnetic spheroids.
\newblock {\em Journal of Applied Physics}, 29(3):318--323, 1958.

\bibitem{doi:10.1063/1.1735216}
P.~C. Fletcher and R.~O. Bell.
\newblock Ferrimagnetic resonance modes in spheres.
\newblock {\em Journal of Applied Physics}, 30(5):687--698, 1959.

\bibitem{princepyig}
Andrew Princep, Russell Ewings, Simon Ward, Sandor Tóth, C.~Dubs, Dharmalingam
  Prabhakaran, and Andrew Boothroyd.
\newblock The full magnon spectrum of yttrium iron garnet.
\newblock {\em npj Quantum Materials}, 2, 12 2017.

\bibitem{Pozar:882338}
David~M Pozar.
\newblock {\em {Microwave engineering; 3rd ed.}}
\newblock Wiley, Hoboken, NJ, 2005.

\bibitem{PhysRev.110.1311}
R.~C. LeCraw, E.~G. Spencer, and C.~S. Porter.
\newblock Ferromagnetic resonance line width in yttrium iron garnet single
  crystals.
\newblock {\em Phys. Rev.}, 110:1311--1313, Jun 1958.

\bibitem{PhysRevLett.3.32}
E.~G. Spencer, R.~C. LeCraw, and A.~M. Clogston.
\newblock Low-temperature line-width maximum in yttrium iron garnet.
\newblock {\em Phys. Rev. Lett.}, 3:32--33, Jul 1959.

\bibitem{yigline}
R~T.~Farrar.
\newblock Spin‐lattice relaxation time in yttrium iron garnet.
\newblock {\em Journal of Applied Physics}, 29, 03 1958.

\bibitem{PhysRevB.48.16407}
Zhang Zheng-Wu, Wu~Ping-Feng, Yang Jian-Hua, and Zhou Kang-Wei.
\newblock Absorption spectrum and zero-field splitting of
  ${\mathrm{y}}_{3}$${\mathrm{fe}}_{5}$${\mathrm{o}}_{12}$.
\newblock {\em Phys. Rev. B}, 48:16407--16409, Dec 1993.

\bibitem{Krupka_1999}
Jerzy Krupka, Stephen~A Gabelich, Krzysztof Derzakowski, and Brian~M Pierce.
\newblock Comparison of split post dielectric resonator and ferrite disc
  resonator techniques for microwave permittivity measurements of
  polycrystalline yttrium iron garnet.
\newblock {\em Measurement Science and Technology}, 10(11):1004--1008, sep
  1999.

\bibitem{1325655}
M.~N. {Afsar}, K.~M. {Lee}, {Yong Wang}, and K.~{Kocharyan}.
\newblock Measurements of complex permittivity and permeability of common
  ferrimagnets at millimeter waves.
\newblock {\em IEEE Transactions on Magnetics}, 40(4):2826--2828, 2004.

\bibitem{doi:10.1063/1.351846}
A.~M. Hofmeister and K.~R. Campbell.
\newblock Infrared spectroscopy of yttrium aluminum, yttrium gallium, and
  yttrium iron garnets.
\newblock {\em Journal of Applied Physics}, 72(2):638--646, 1992.

\bibitem{doi:10.1063/1.3626057}
Takanori Tsutaoka, Teruhiro Kasagi, and Kenichi Hatakeyama.
\newblock Permeability spectra of yttrium iron garnet and its granular
  composite materials under dc magnetic field.
\newblock {\em Journal of Applied Physics}, 110(5):053909, 2011.

\bibitem{PhysRev.134.A1581}
Elmer~E. Anderson.
\newblock Molecular field model and the magnetization of yig.
\newblock {\em Phys. Rev.}, 134:A1581--A1585, Jun 1964.

\bibitem{PhysRevLett.124.171801}
N.~Crescini, D.~Alesini, C.~Braggio, G.~Carugno, D.~D'Agostino,
  D.~Di~Gioacchino, P.~Falferi, U.~Gambardella, C.~Gatti, G.~Iannone, C.~Ligi,
  A.~Lombardi, A.~Ortolan, R.~Pengo, G.~Ruoso, and L.~Taffarello.
\newblock Axion search with a quantum-limited ferromagnetic haloscope.
\newblock {\em Phys. Rev. Lett.}, 124:171801, May 2020.

\bibitem{inprep}
N.~Crescini et~al.
\newblock Studying the dynamics of photon-magnon hybrid systems with ultrafast
  laser pulses.

\bibitem{PhysRevB.95.214423}
H.~Maier-Flaig, S.~Klingler, C.~Dubs, O.~Surzhenko, R.~Gross, M.~Weiler,
  H.~Huebl, and S.~T.~B. Goennenwein.
\newblock Temperature-dependent magnetic damping of yttrium iron garnet
  spheres.
\newblock {\em Phys. Rev. B}, 95:214423, Jun 2017.

\end{thebibliography}


\begin{appendix}
\section{Langevin equations and the effective Hamiltonian}
\label{app:matrix}
The dynamics of the system can be described by the quantum Langevin equations
\begin{align}
\begin{split}
\dot{x}&=-i(\omega_x-i\gamma_x)x-i\sum_{y\in Y} \frac{g_{xy}}{2}y \\
\dot{y}&=-i(\omega_y-i\gamma_y)y-i\sum_{x\in X} \frac{g_{xy}}{2}x - i\sum_{y*\neq y} \frac{g_{y y*}}{2}y*,
\label{eq:aLang}
\end{split}
\end{align}
which are valid for every cavity mode $x\in X$ and magnon mode $y\in Y$, and where $i$ is the imaginary unit and $y*\in Y$. For the sake of this work it suffices considering only the intrinsic decay rates of the cavity modes, as they dominate over the ones introduced by external couplings if we assume that the setup's antennas are weakly coupled.
Eq.s\,(\ref{eq:aLang}) may be recast in their matrix form as
\begin{equation}
\dot{W}=-i{\cal H}W,
\label{eq:lang_mat}
\end{equation}
where $W=\{X,Y\}=\{ x_1, x_2, \dots, x_N,y_1, y_2, \dots, y_M \}$, and ${\cal H}$ is the matrix reported in Eq.\,(\ref{eq:Heff}) and used throughout this work, in particular to determine $s_{x_1}(\omega,\omega_m)$ in Eq.\,(\ref{eq:s}). Here we note that it is experimentally challenging to have an antenna coupled to a single mode, making our calculated transmission spectra an incomplete description of the apparatus. Nevertheless, as is detailed in Section \ref{sec:er} and Fig.\,\ref{fig:tm110}, this approximation holds thanks to the geometric configuration of the two considered cavity modes.
We stress that one can get a more realistic description of the HS spectra by summing over the possible transmissions due to external couplings to other modes as
\begin{equation}
s_\mathrm{tot}(\omega,\omega_m)=\sum_{x\in X} \eta_x s_x(\omega,\omega_m),
\label{eq:stot}
\end{equation}
where $\eta_x$ is a constant accounting for the coupling strength between the antenna and the different cavity modes.
This is an effective approximation which resists if all the external couplings are weak, and do not influence the linewidths of the modes. An even better description consists in the addition of a mode-dependent dissipation in Langevin equations to account for the coupling of the antennas, see for example Ref.\,[\onlinecite{PhysRevB.99.054404}], which naturally yields the increase of the modes' decay rate.


\section{Spheres production}
\label{app:tumbling}

\begin{figure}
	\centering
	\includegraphics[width=.48\textwidth]{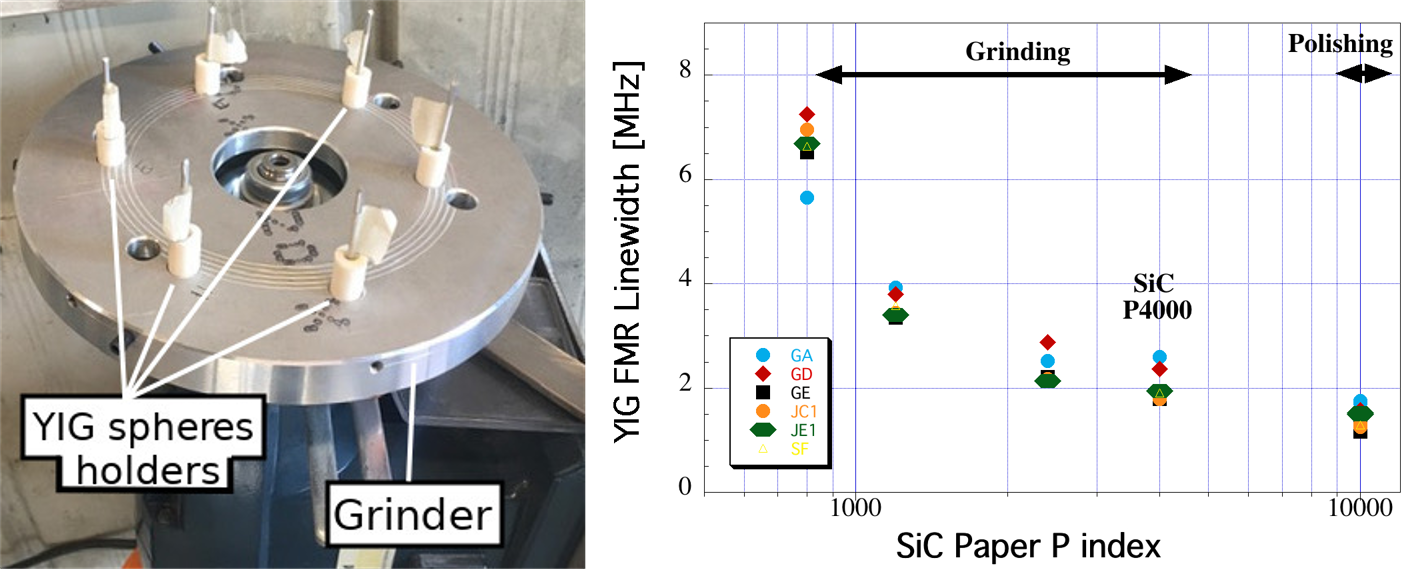}
	\caption{The instrument used to produce the spheres (left). Improvement of linewidth with subsequent steps for a batch of six spheres (right).}
	\label{fig:flessibile}
	\end{figure}

Super polished YIG spheres show the sharpest magnetic resonance linewidth\cite{PhysRev.110.1311,PhysRevLett.3.32,yigline}. Material purity and surface roughness are the two key elements to be cured in order to obtain the best linewidth values at all temperatures\cite{PhysRevB.95.214423}. Our spheres have been produced on site to have better control of all the relevant parameters. We buy large single crystal high purity YIG cylinders, normally a few cm length and 5 to 7 mm diameter, and cut them into 3\,mm size cubes. Each cube then follows a grinding procedure to obtain spheres of about 2.1 mm diameter. The grinder is based on a high frequency rotating plate with replaceable silicon carbide (SiC) sandpaper foils where the YIG piece get abraded while tumbling inside a plastic holder (see the left plot in Fig.\,\ref{fig:flessibile}). Starting from a P800 SiC paper, finer and finer grit size are used in sequence up to the final P4000. The duration of each step has been optimized after several trials. The last step consists of four passages with P4000 virgin SiC paper, each lasting half an hour. At the end of this process, we get spheres with the nominal diameter within a few percent and linewidths of about 2\,MHz. A final 24 hours polishing with Alumina-based suspension on a low frequency rotating system results in a linewidth slightly above 1.3\,MHz at room temperature (see the right plot of Fig.\,\ref{fig:flessibile}).

\section{Hot-bore superconducting magnet and millikelvin temperature hybrid system}
\label{app:apparatus}
The setup used for testing the single and multiple-spheres configurations includes a superconducting magnet, described in Section \ref{sec:er}, with a hot bore in which the HS is inserted.
This custom setup was built to test several HSs per-day, which is incompatible with multiple refrigeration cycles. The cavity is mounted on a plastic support that aligns its axis with the one of the magnet, and a PTFE support holds the pipe which positions the YIG spheres on this same axis.

At milli-Kelvin temperatures, the cavity is anchored to the mixing chamber of a dilution refrigerator with two large copper bars. Its temperature is monitored with a thermometer attached to the cavity body. The setup must ensure a proper thermalization of the YIG spheres, which for this reason are contained in a pipe filled with gaseous helium. The preparation of the fused silica pipe is done at room temperature as follows. A vacuum system is designed in such a way to remove the air from the pipe, which is then immersed in a 1 bar helium controlled atmosphere and closed. In this way the pipe is filled with helium, and can be sealed by using a copper plug glued with Stycast. First the sealing is tested without the samples by measuring the frequency shift of the TM110 mode of the cavity-pipe system with and without helium. The frequency is measured with the helium-filled pipe, which is then immersed in liquid nitrogen and again placed in the cavity. Re-measuring the same frequency excludes the presence of leaks. The plug used to close the pipe is eventually anchored to the cavity body with a copper rod, granting a good thermalization of the whole system.

\end{appendix}

\end{document}